\documentclass[twocolumn]{aastex631}

\usepackage{amsmath}
\usepackage{CJK}

\shorttitle{Polarization in scattering by aligned grains}
\shortauthors{Yang and Li}
\graphicspath{{./}{figures/}}
\begin{document}
\begin{CJK*}{UTF8}{gbsn}

\title{Probing Magnetic Fields in Protoplanetary Disk Atmospheres through Polarized near-IR Light Scattered by Aligned Grains} 

\correspondingauthor{Haifeng Yang}
\email{hfyang@pku.edu.cn}

\author[0000-0002-8537-6669]{Haifeng Yang (杨海峰)}
\altaffiliation{Boya Fellow}
\affil{Kavli Institute for Astronomy and Astrophysics, Peking University, Yi He Yuan Lu 5, Haidian Qu, Beijing 100871, People's Republic of China}

\author[0000-0002-7402-6487]{Zhi-Yun Li}
\affiliation{Department of Astronomy, University of Virginia, Charlottesville, VA 22903, USA}



\begin{abstract}

Magnetic fields play essential roles in protoplanetary disks.
Magnetic fields in the disk atmosphere are of particular interest, as they are connected to the wind-launching mechanism.
In this work, we study the polarization of the light scattered off of magnetically aligned grains in the disk atmosphere, focusing on the deviation of the polarization orientation from the canonical azimuthal direction, which may be detectable in near-IR polarimetry with instruments such as VLT/SPHERE.
We show with a simple disk model that the polarization can even be oriented along the radial (rather than azimuthal) direction, especially in highly inclined disks with toroidally dominated magnetic fields. This polarization reversal is caused by the anisotropy in the polarizibility of aligned grains and is thus a telltale sign of such grains.
We show that the near-IR light is scattered mostly by $\mu$m-sized grains or smaller at the $\tau=1$ surface and such grains can be magnetically aligned if they contain superparamagnetic inclusions.
For comparison with observations, we generate synthetic maps of the ratios of $U_\phi/I$ and $Q_\phi/I$, which can be used to infer the existence of (magnetically) aligned grains through a negative $Q_\phi$ (polarization reversal) and/or a significant level of $U_\phi/I$.
We show that two features observed in the existing data, an asymmetric distribution of $U_\phi$ with respect to the disk minor axis and a spatial distribution of $U_\phi$ that is predominantly positive or negative, are incompatible with scattering by spherical grains in an axisymmetric disk. They provide indirect evidences for scattering by aligned non-spherical grains.

\end{abstract}



\section{Introduction}

The evolution of protoplanetary disks is generally thought to be determined by magnetic fields through either magnetorotational instability \citep{Balbus1991} or magnetized disk wind \citep{Blandford1982}.
The magnetic field structure in the disk atmosphere\footnote{In the context of disk dynamics, the disk atmosphere is often defined as where the disk winds are launched. In the context of near-IR scattering polarimetry, the disk atmosphere is the optical depth of unity surface. Both locations are a few gas scale heights above the disk midplane, and we use ``disk atmosphere" to denote both, even though they are not strictly at the same location.}
is of particular interest in understanding the wind launching mechanisms. Strongly magnetized disks tend to launch magneto-centrifugal wind (MCW; \citealt{Blandford1982}) with rigid and mostly poloidal magnetic field lines in the disk atmosphere. Weakly magnetized disks tend to launch magneto-thermal disk winds \citep{Bai2016}, and rely on the vertical gradient of the magnetic pressure from the toroidal field to launch the wind. 
The magnetic field strength is also related to other interesting issues of protoplanetary disks, such as accretion rates. 

Dust grains can trace the magnetic field if they are magnetically aligned \citep{Andersson2015,Lazarian2007}. 
While polarized (sub)millimeter dust thermal emission has been proven to be a powerful tool to study magnetic fields on scales larger than the disks (e.g., \citealt{Planck2018XII}; \citealt{HullZhang2019} and references therein), its application on the disk scale has not been as successful because we see mostly scattering-induced polarization at shorter (e.g., $870\rm\, \mu m$)  wavelengths \citep{Stephens2017,Hull2018,Bacciotti2018,Dent2019} and complicated non-magnetic-origin patterns at longer (e.g., $3$ mm) wavelengths \citep{Kataoka2017,Harrison2019}. \cite{Yang2021} showed that the Larmor precession in the disk midplane is likely too slow to ensure magnetic alignment, which is likely the reason behind the failure of polarized dust thermal emission to trace magnetic fields in disks. 
However, \cite{Yang2021} proposed that the micron-sized dust grains in the disk atmosphere can potentially be magnetically aligned. 
This idea was partially supported by \cite{Li2016}, who found polarized radiation at $10.3\rm\, \mu m$, which may be explained in part by thermal emission from grains aligned with the magnetic field in the disk atmosphere. 

The Spectro-Polarimetric High-contrast Exoplanet REsearch (SPHERE) at the Very Large Telescope (VLT) has been used for high-resolution polarimetric observations of protoplanetary disks in scattered near-IR light \citep{Benisty2015,Avenhaus2018,Garufi2020,Garufi2022}. 
These studies usually focus on the azimuthal Stokes parameter $Q_\phi$ component to produce high resolution images of protoplanetary disks. The $U_\phi$ component is often observed to be small and its divergence from zero is a sign of deviation from the simplest single Rayleigh scattering.
%An exception is 
Theoretically, \cite{Canovas2015} used a generic transition disk model with large dust grains to show that the scattered light from moderately inclined disks can possess a significant $U_\phi$. 
\cite{Whitney2002} studied the scattering by aligned grains, but focused on the circular polarization in the protostellar envelope. Our focus in this paper is on the near-IR photons scattered by aligned dust grains in the atmosphere of protoplanetary disks, which can potentially have polarization patterns different from the commonly expected pure azimuthal ones. 

The structure of the paper is as follows. 
In Section~\ref{sec:basic} and Appendix~\ref{sec:GF}, we discuss the polarization of scattered light in the grain's frame, focusing on the difference with spherical dust grains. In Section~\ref{sec:DF}, we calculate the polarization pattern in a disk configuration. In Section~\ref{sec:discussion}, we discuss the grain size distribution and grain alignment at the optical depth $\tau=1$ surface in a generic protoplanetary disk model. In Section~\ref{sec:added_discussion}, we discuss our results, including detectability of the $U_\phi$ produced by the scattering of magnetically aligned grains, the differences between the patterns produced by aligned grains and multiple scattering of spherical grains, and implications for observations. We summarize our results in Section~\ref{sec:summary}.

\section{Basic physics}
\label{sec:basic}
\begin{figure}
    \centering
    \includegraphics[width=0.4\textwidth]{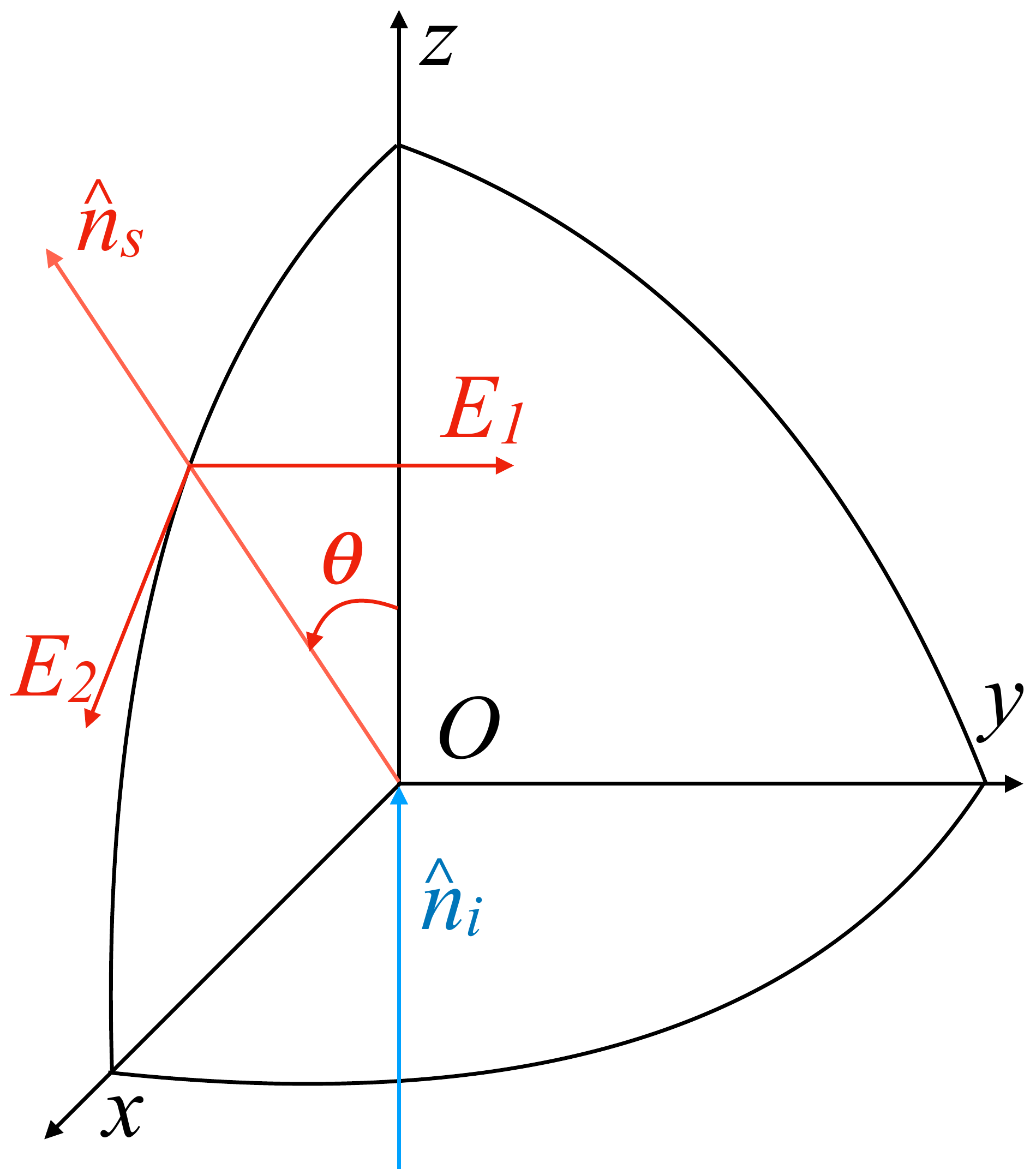}
    \caption{The geometry of a simple setting in the grain's frame. The photon propagating in the $\hat{z}$ direction is scattered by the dust grain sitting at the origin $O$ towards the $\hat{n}_s$ 
    direction in the $xOz$ plane. The scattering angle is $\theta$. The scattered light is decomposed into $E_1$ and $E_2$ directions. Note that it is often assumed that the scattered light is polarized along $E_1$ direction.}
    \label{fig:physics}
\end{figure}

Before considering the general case in a disk environment. we first illustrate the basic physics of why the polarization orientation of the light scattered by aligned grains can be different from that by spherical grains in a simple setup in the grain's frame.  
We consider a Cartesian coordinate system $xyz$ (see Figure~\ref{fig:physics}). Let $\hat{z}$ be the propagation direction for incoming light. Without loss of generality, we fix the scattered light in the $xOz$ plane, with $O$ being the particle location. 
Let $\theta$ be the scattering angle, the scattering directional vector is then $(\sin\theta,0,\cos\theta)$. Since light is transverse wave, its $E$ vector can be decomposed into two components that are perpendicular to the scattering direction.
We call the component perpendicular to both incoming and scattering lights $E_1$ (in the $\hat{e}_1$ direction), and the other component $E_2$ (in the $\hat{e}_2$ direction). We can easily see that the $E_1$ direction is also the $\hat{y}$ direction of our coordinate system.
If the dust particle is spherical, the scattered light can either be polarized along $E_1$ or along $E_2$ direction, due to the symmetry of this scattering geometry. 
If we define the Stokes parameters such that fully polarized light with polarization along $E_1$ has $Q=I$, the Stokes $U$ is always zero for spherical dust grains.
If the dust particle is not spherical, and if $\hat{x}$ and $\hat{y}$ are not the principle axes of the dust particle, the light will no longer be polarized along either $E_1$ or $E_2$ direction, which leads to non-zero Stokes $U$ component.
The break of the symmetry is the reason why the polarization can deviate from the spherical case. 

It turns out that the deviation is always maximized in the forward and backward scattering directions. For the forward scattering, we have the scattering direction being the same as the incoming light direction ($\hat{z}$). If the dust particle is spherical,
the scattered light will always be non-polarized, because of the symmetry. If the dust particle is non-spherical, say being elongated along $\hat{x}$ direction, then the scattered light would be polarized along the $\hat{x}$ direction, which is qualitatively different from the spherical case. 
If we change the scattered light slightly away from the forward scattering direction with a small scattering angle $\theta$ in the $xOz$ plane, then the light scattered by a spherical dust grain would be polarized along $\hat{y}$ direction, in the often assumed Rayleigh scattering regime. 
For dust grains elongated along $\hat{x}$ direction, on the contrary, this small deviation in scattering angle is not enough to change the polarization state of the scattered light, and the scattered light is still polarized along $\hat{x}$ direction. 
Hence \textit{the angle difference} between polarization orientation of light scattered by spherical dust grains and the polarization orientation of light scattered by elongated dust grains near the forward scattering direction \textit{can always be as large as $90^\circ$}, 
as they are perpendicular to each other in the set-up we discussed above. 

In Appendix~\ref{sec:GF}, we discuss the angle difference in the grain's frame in more detail. We show that the angle difference between the spherical dust grain and the elongated dust grain can easily reach $10^\circ$ in the grain's frame.
We also show that the angle difference increases with the angle between the symmetry axis of the dust particle and the incoming light direction and with the aspect ratio of the dust particle. 
It also depend on the compositions of the dust grains. 
In what follows, we focus on the angle difference in a disk environment where most of the relevant observations are carried out. 

\section{Polarization in the disk atmosphere}
\label{sec:DF}
\subsection{Model prescription}

Calculations in Section~\ref{sec:basic} and Appendix~\ref{sec:GF} focus on the polarization of the scattered light in the grain's frame. While more physically intuitive, it is not directly connected to the observed polarization. Here we study the scattered light in a protoplanetary disk, focusing on the deviation of polarization orientation from the direction perpendicular to the stellar light, the expected polarization orientation in the small spherical particle regime. 

\begin{figure}
    \centering
    \includegraphics[width=0.45\textwidth]{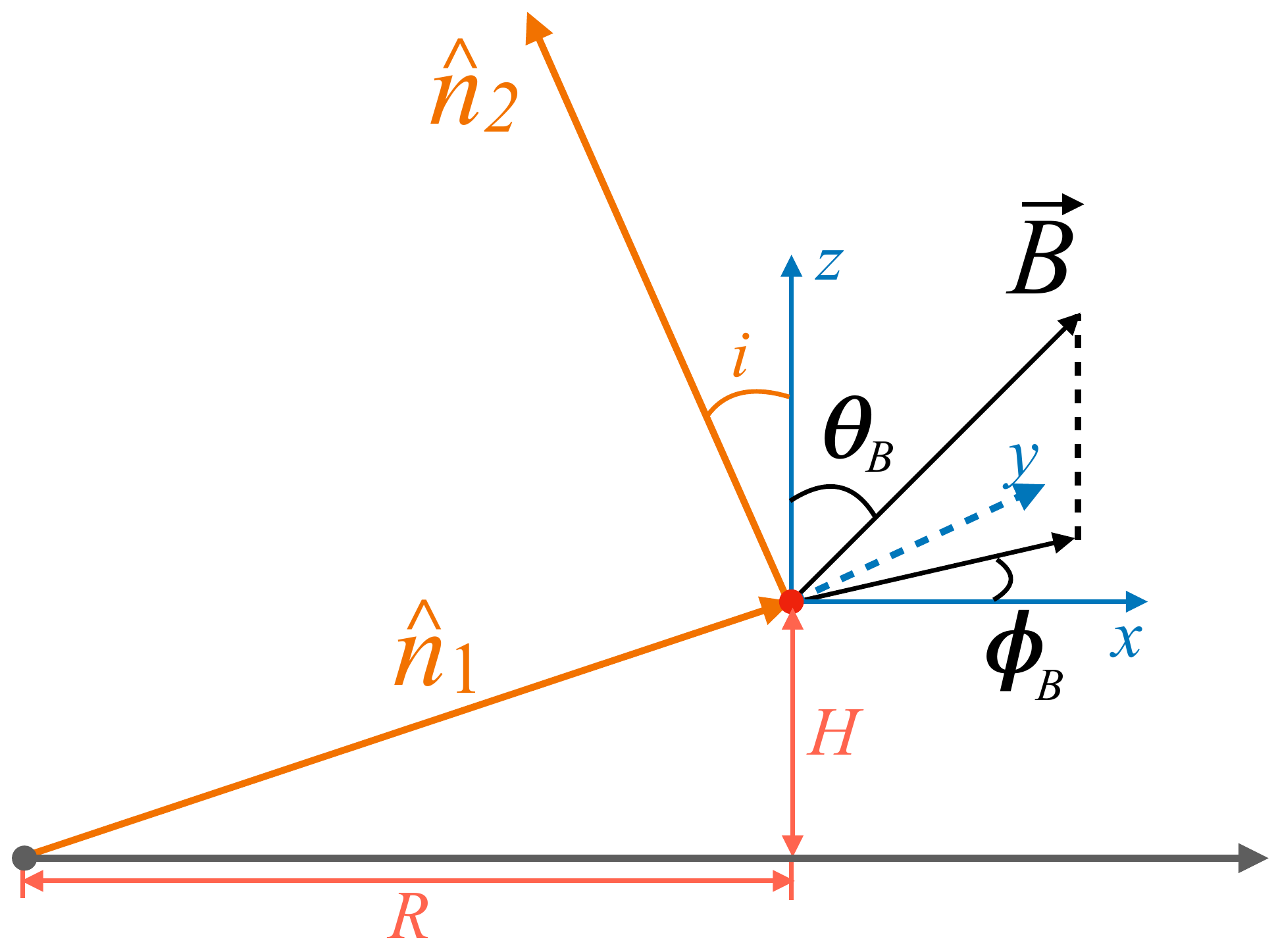}
    \caption{The geometry of our set-up in the disk frame. The black dot represents the central star. The red dot represents the scattering dust grain. Hence $\hat{n}_1$ connecting these two dots is the incoming light direction. The scattered light propagates along $\hat{n}_2$, making an angle of $i$ with the $z$ axis. The local magnetic field direction is prescribed by the angles $\theta_B$ and $\phi_B$. See text for more details.}
    \label{fig:DF}
\end{figure}

To calculate the polarization orientation in the scattered light from the surface of a protoplanetary disk, we consider a set-up shown in Figure~\ref{fig:DF}. The black horizontal arrow represents the disk midplane. The red dot represents the dust particle that scatters light from $\hat{n}_1$ direction towards the $\hat{n}_2$ direction, which makes an angle $i$ with the $z$ direction, the direction perpendicular to the disk midplane; $i$ is simply the disk inclination angle ($i=0$ is face-on). Note that only one dust grain is plotted in the figure, but it represents a ring of dust grains all of which have the same cylindrical radius $R$ from the star and height $H$ above the disk midplane. We assume that the local magnetic field $\mathbf{B}$ makes an angle $\theta_B$ with the $z$ direction, and an azimuthal angle $\phi_B$ from the $x$ direction, such that $\phi_B=0$ implies a pure poloidal magnetic field. As usual, we assume that the grains are aligned with their shortest axis along the magnetic field direction. If the grains are spinning around the B field, they would be effectively oblate after assemble-averaging independent of their intrinsic shapes. 

We adopt the same dust composition as used in Appendix~\ref{sec:GF}, which is the same as the one adopted by \cite{Birnstiel2018}, and assume the dipole approximation for simplicity. We consider $5$ parameters: $H/R$, $i$, $\theta_B$, $\phi_B$, and the dust aspect ratio $s$. 

\subsection{A limiting case: purely toroidal magnetic field}
\label{ssec:toroidal}

\begin{figure}
\includegraphics[width=0.45\textwidth]{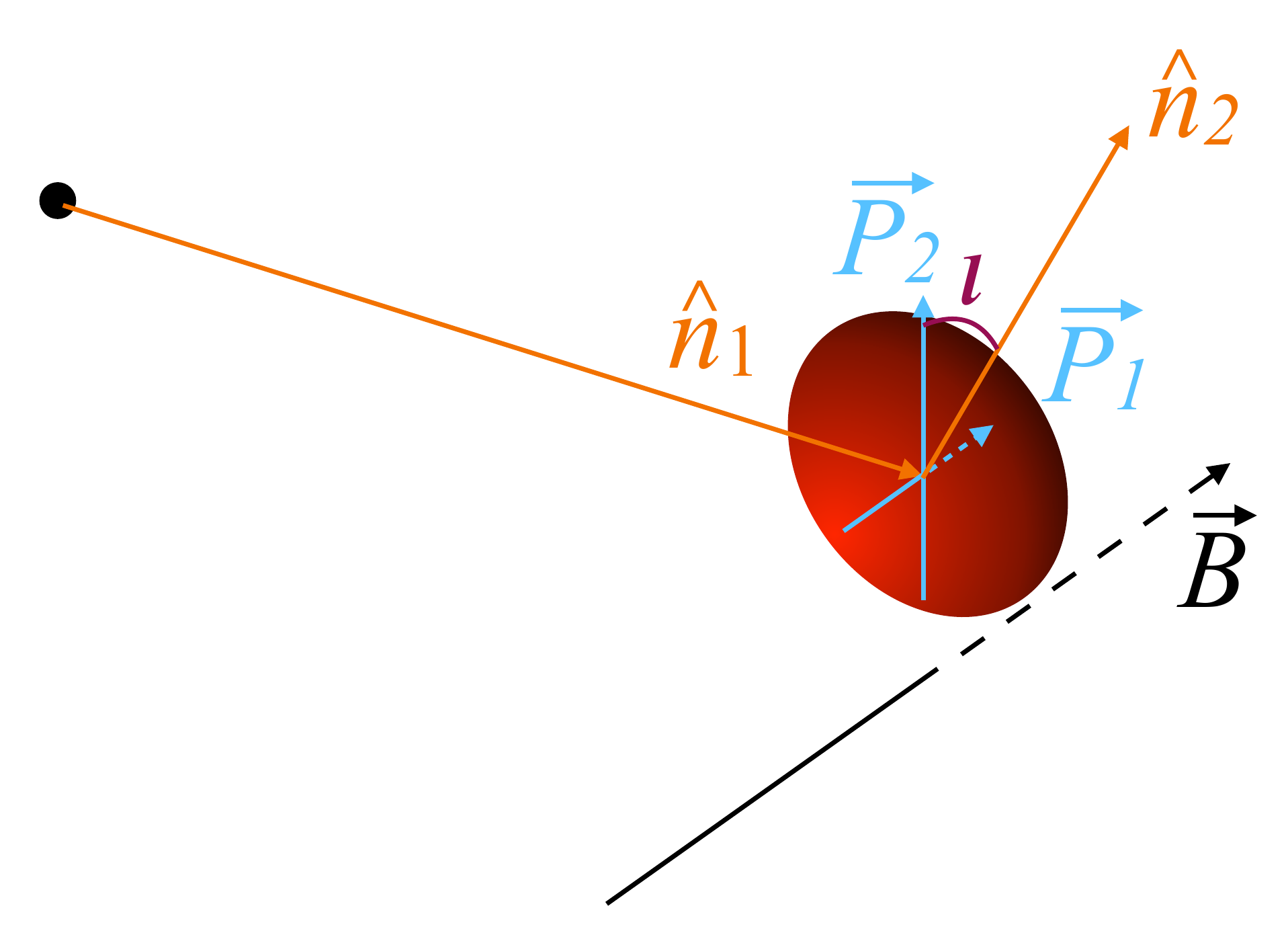}
\caption{The geometry of an oblate grain aligned by a pure toroidal magnetic field. By definition, the magnetic field $\mathbf{B}$ is perpendicular to incoming light direction $\hat{n}_1$. The incoming light excites two dipoles in the grain, denoted as $\mathbf{P}_1$ and $\mathbf{P}_2$. The scattered light direction $\hat{n}_2$ makes an angle of $\iota$ with $\mathbf{P}_2$.
Note that we have $|\mathbf{P}_2|>|\mathbf{P}_1|$ under this configuration.}
\label{fig:toroidal}
\end{figure}

Before studying the polarization pattern in a generic model, we will first consider a limiting case that will help our understanding: a purely toroidal magnteic field along $\hat{y}$ direction, with $\theta_B=90^\circ$ and $\phi_B=90^\circ$. The incoming light propagating along $\hat{n}_1$ direction can be decomposed along two directions: $\hat{y}$ and $\hat{n}_1\times \hat{y}$. We will denote the dipoles excited by these two components $\mathbf{P}_1$ and $\mathbf{P}_2$, respectively. 
See Figure~\ref{fig:toroidal} for a schematic illustration of this setting.

The angle between the scattered light direction $\hat{n}_2$ and the $\mathbf{P}_2$, defined as $\iota$, will be important to our analysis that follows. Note that in the disk frame, we have $\iota=i +(-) \mathrm{tan}^{-1}(H/R)$ for the near (far) side of the ring considered in Figure~\ref{fig:DF}. 

\begin{figure}
    \centering
    \includegraphics[width=0.45\textwidth]{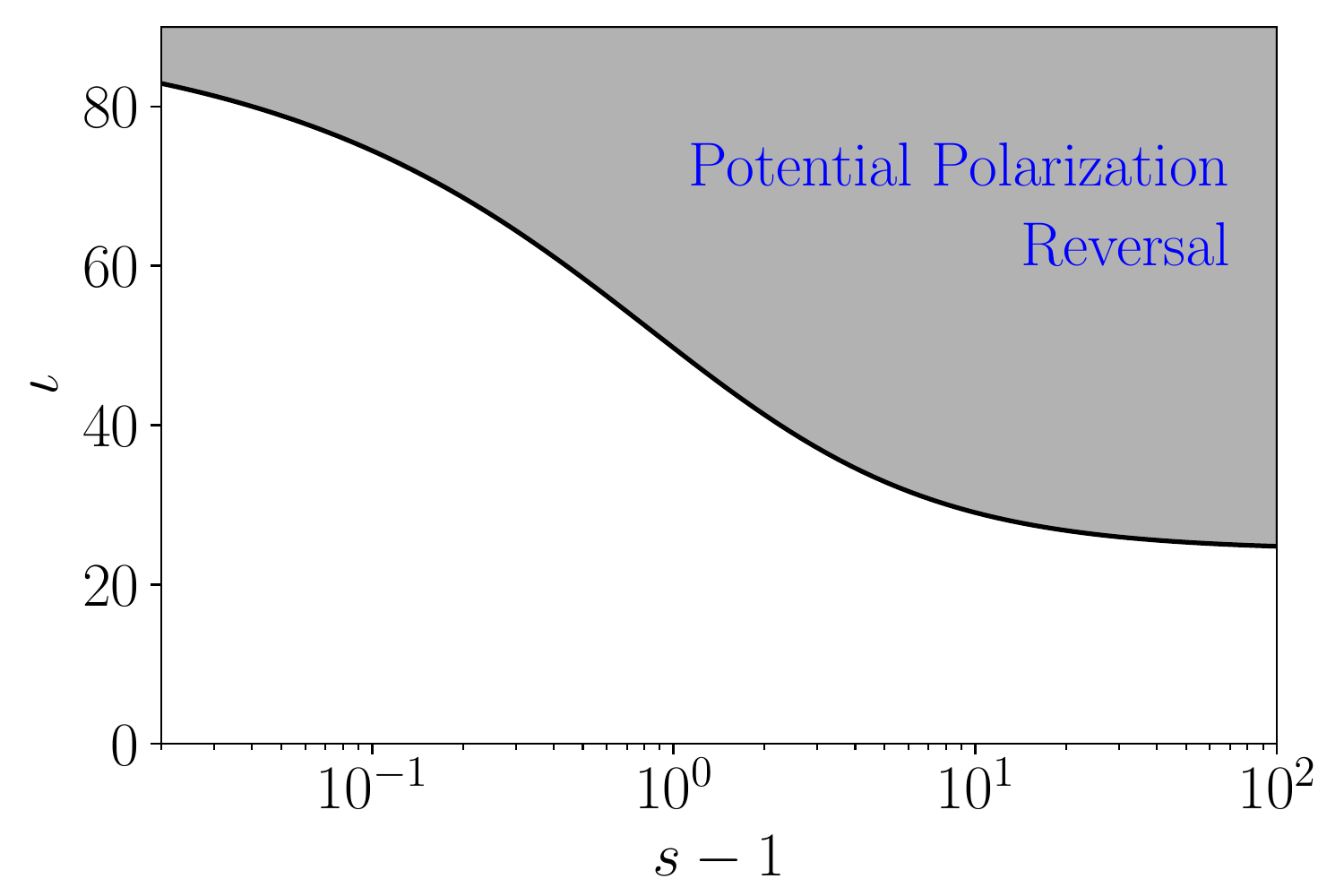}
    \caption{Critical curve for polarization reversal. 
    The $s$ in the x-axis is the aspect ratio of dust grains. The $\iota$ is the angle between the scattered light propagating direction $\hat{n}_2$ and the second dipole direction $\mathbf{P}_2$ (c.f. Figure~\ref{fig:toroidal}). 
    In the gray parameter space, the polarization is potentially reversed, i.e. the light is polarized along that radial, rather than the canonical azimuthal direction. }
    \label{fig:iotac}
\end{figure}

Because the dust grains are aligned with the magnetic field in $\mathbf{P}_1$ direction, we have $P_1=\alpha_3 E_1$. Since $\mathbf{P}_1$ is always perpendicular to $\hat{n}_2$, we have $E_{s1}\propto P_1\propto \alpha_3$, where $E_{s1,2}$ is the $E$ vector of scattered light induced by $\mathbf{P}_{1,2}$. Similarly, we have $P_2=\alpha_1 E_2$. Since $\mathbf{P}_2$ is making an angle $\iota$ with $\hat{n}_2$, we have $E_{s2}\propto P_2\sin \iota\propto \alpha_1\sin\iota$. If $E_{s1}>E_{s2}$, the scattered light is polarized along $\mathbf{E}_{s1}$ direction, i.e. the direction perpendicular to both $\hat{n}_1$ and $\hat{n}_2$, which is the canonical (azimuthal) polarization direction for spherical particles. If $E_{s1}<E_{s2}$, the scattered light is polarized along $\mathbf{E}_{s2}$ direction, i.e. the $\mathbf{P}_2$ projected towards the sky plane, the plane perpendicular to $\hat{n}_2$. This is also the direction of the projected $\hat{n}_1$ (radial) direction and is perpendicular to canonical polarization direction for spherical particles. The polarization direction is changed by $90^\circ$ and the $Q_\phi$, the azimuthal $Q$ parameter \citep{deBoer2020}, changes from positive to negative. We call this ``polarization reversal".
Define the critical angle $\iota_c$ as $\iota_c\equiv \sin^{-1}(|\alpha_3|/|\alpha_1|)$. For $\iota>\iota_c$ ($\iota<\iota_c$), we have $E_{s2}>E_{s1}$ ($E_{s2}<E_{s1}$), and the polarization is (not) reversed. The dependence of the critical angle with dust aspect ratio is shown in Figure~\ref{fig:iotac}. 

In Section~\ref{ssec:fidDF}, we discuss a fiducial case of a moderately inclined disk with $\iota>\iota_c$ but not a purely toroidal magnetic field. We will demonstrate that the polarization reversal still exists (i.e., it is not limited to the pure toroidal magnetic field configuration) and is likely once we reach the critical angle. In Section~\ref{ssec:DFmodel2}, we present a less inclined disk model without polarization reversal. We will show that there is still an appreciable angle deviation from the azimuthal pattern that is potentially detectable with VLT/SPHERE.

\subsection{Fiducial case}
\label{ssec:fidDF}

\begin{figure*}
    \centering
    \includegraphics[width=\textwidth]{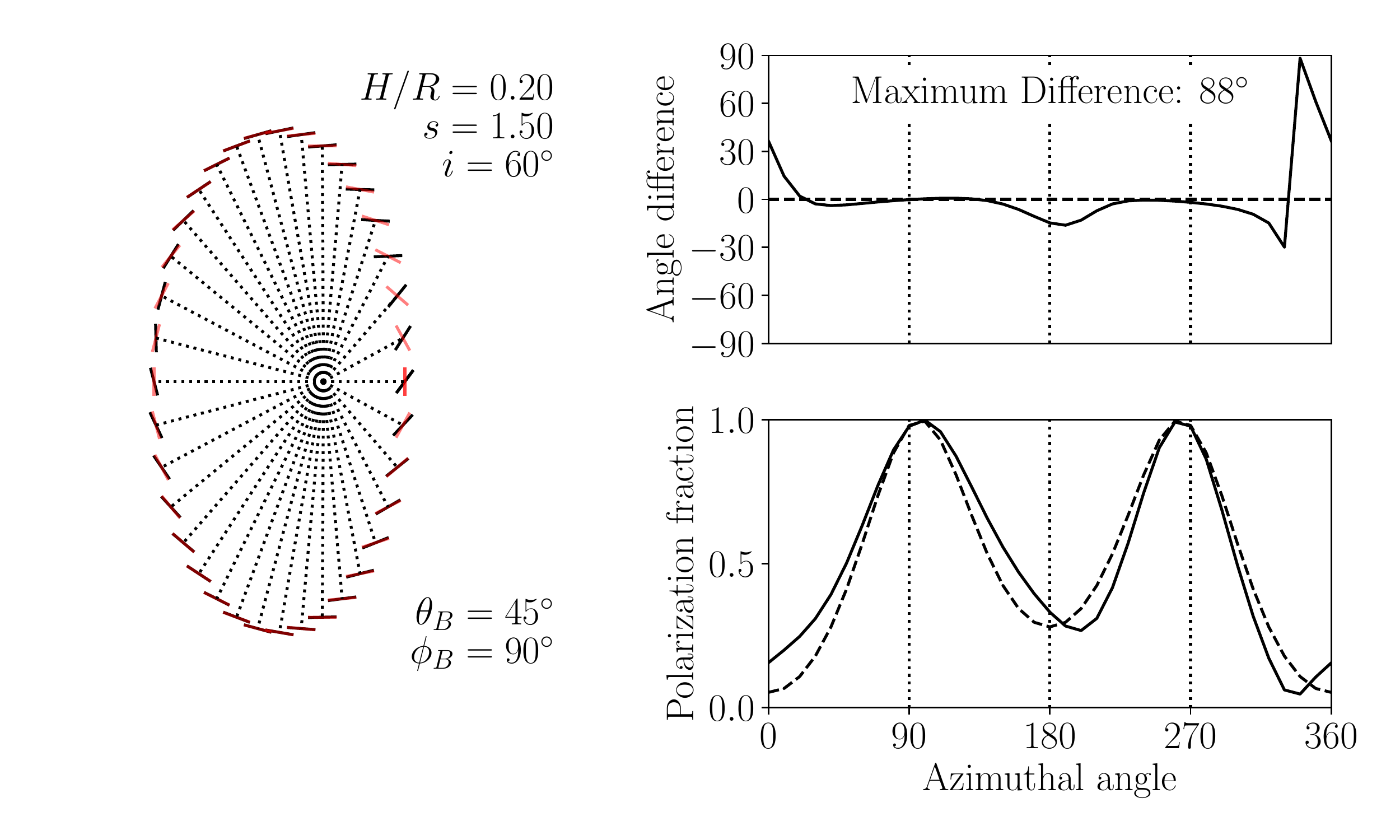}
    \caption{The polarization pattern in our fiducial disk model (Model 1). The parameters adopted are listed in the left panel. See Figure~\ref{fig:DF} for their definitions. 
    \textit{Left:} The black uni-length line segments represent the polarization orientation. 
    The dotted lines connect the central star with each test scattering grain. 
    The red line segments are perpendicular to the dotted lines, and represent the often assumed azimuthal polarization patterns.
    \textit{Upper right:} The difference in polarization orientation as a function of the azimuthal angle of the test scattering grain in the disk frame. The polarization is essentially ``reversed" and is along the radial direction at an azimuthal angle of $330^\circ$. 
    \textit{Lower right:} The polarization fraction as a function of the azimuthal angle. The solid line represents the fiducial case, whereas the dashed line represents the small spherical grain case.}
    \label{fig:disk_fid}
\end{figure*}

For the fiducial case, we consider $H/R=0.2$, $i=60^\circ$, $s=1.5$, $\theta_B=45^\circ$, and $\phi_B=90^\circ$ (Model 1 hereafter). 
The last two angles imply that the toroidal to poloidal magnetic field ratio is $B_\mathrm{tor}/B_\mathrm{pol}=1$. 
We can easily calculate the $\iota$ at the near side as $\iota=i+\tan^{-1}(H/R)=71^\circ>\iota_c=59^\circ$.
The results are shown in Figure~\ref{fig:disk_fid}. In the left panel, we show the polarization orientation at each location on a dust ring of constant cylindrical radius $R$ and height $H$ in the sky plane, with black uni-length line segments. To guide the eyes, we also use dotted lines to connect the central star and the scattering grains. In the often assumed case, the polarization is perpendicular to the dotted line, shown as red lines, with only the $Q_\phi$ component being non-zero. 
We can see that the polarization is completely reversed near azimuthal angle of $330^\circ$. 
To show the deviation more quantitatively, we plot the angle difference as a function of the azimuthal angle (of dust grains in the disk frame) in the upper right panel. The polarization fraction, the ratio $p\equiv \mathrm{PI}/\mathrm{I}$ between the polarized intensity and the total intensity, as a function of the azimuthal angle is shown in the lower right panel, as a solid line. The polarization fraction for scattering by small spherical grains is also shown as a dashed line. We can see that the polarization is still maximized at $100\%$, while the phase function deviates from the spherical curve slightly. 
We can see that the polarization reversal location coincides with low polarization points, because it relies on the usually sub-dominant component $\mathbf{P}_2$ to overwhelm the $\mathbf{P}_1$ to have polarization reversal (see Section~\ref{ssec:toroidal}).

\begin{figure}
    \centering
    \includegraphics[width=0.5\textwidth]{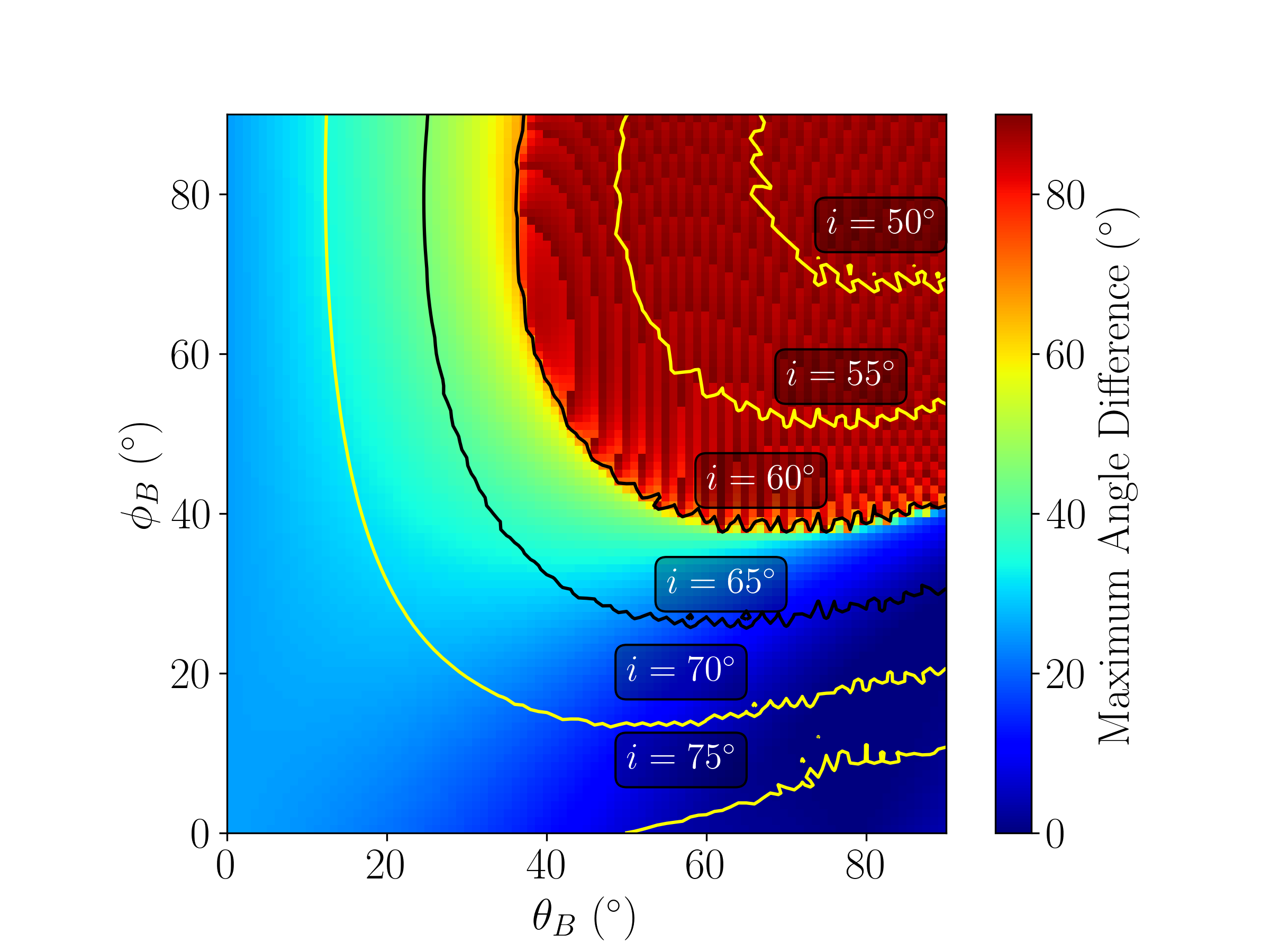}
    \caption{The maximum angle difference for different configurations of magnetic fields, assuming an inclination angle of $60^\circ$, $H/R$ of $0.2$, and a dust aspect ratio of $s=1.5$. The color map represent the maximum angle difference $\Delta\eta$. Also plotted are contours of constant angle difference $\Delta\eta=70^\circ$ for different inclination angles. The inclination angle $i$ is labeled next to each line. }
    \label{fig:Bphase}
\end{figure}

The same calculation is repeated on a finer azimuthal grid ($360$ points to sample the full azimuthal extent) with different combinations of the angles $\theta_B$ and $\phi_B$ that specify the magnetic field configurations. The maximum angle difference for each combination is shown as a colormap in Figure~\ref{fig:Bphase}, with $\theta_B$ and $\phi_B$ as the $x$ and $y$ axis, respectively. 
We can see that the $\phi_B$ has a very strong effect on the angle difference. At the upper right part of the Figure~\ref{fig:Bphase}, the maximum angle difference is essentially $90^\circ$ (i.e., polarization reversal). The fluctuations are due to the finite resolution of the azimuthal grid, and we have tested that the fluctuations become smaller (with the angle deviation closer to $90^\circ$) with an increasing number of grid points. 
Away from the upper right polarization reversal region, the angle difference is still appreciable, and can easily be above $20^\circ$ for our fiducial disk inclination of $60^\circ$. 

To study the dependence of the polarization reversal region on the observing inclination angle $i$, we first find for the fiducial $i=60^\circ$ all pairs of $\theta_B$ and $\phi_B$ that would make the angle difference $\Delta\eta=70^\circ$ \footnote{The choice of $\Delta\eta=70^\circ$ is somewhat arbitrary. Because the steep gradient towards the polarization reversal region, other choices between $\sim 70^\circ$ and $\sim 85^\circ$ have little effect on the contours except for the one for the most inclined ($i=75^\circ$) case. We choose $70^\circ$ to minimize the wiggles on the $i=75^\circ$ curve.}. The resulting constant angle difference $\Delta\eta=70^\circ$ contour is labeled in Figure~\ref{fig:Bphase}. We then repeat the same calculation for several inclination angles. The resulting contours are labeled in the figure as well. We can see that the parameter space with polarization reversal as marked by the $\Delta\eta=70^\circ$ contour increases with increasing inclination angle. 
In the most inclined case, the polarization reversal is almost inevitable, with only a small region in the lower right corner being not completely reversed (the maximum angle difference is still large). 
We would like to note that highly inclined disks are also subject to strong forward scattering and potentially multiple scattering \citep{Canovas2015}. These effects, not considered in this simple first study, may change the results substantially.

\subsection{Less inclined case}
\label{ssec:DFmodel2}
We have shown that for highly inclined disks, the polarization can be reversed, with an orientation along the radial, rather than the canonical azimuthal, direction. We now focus on less inclined cases, particularly the dependence of the maximum angle difference with the observing inclination angle $i$ in the disk frame.

\begin{figure*}
    \centering
    \includegraphics[width=\textwidth]{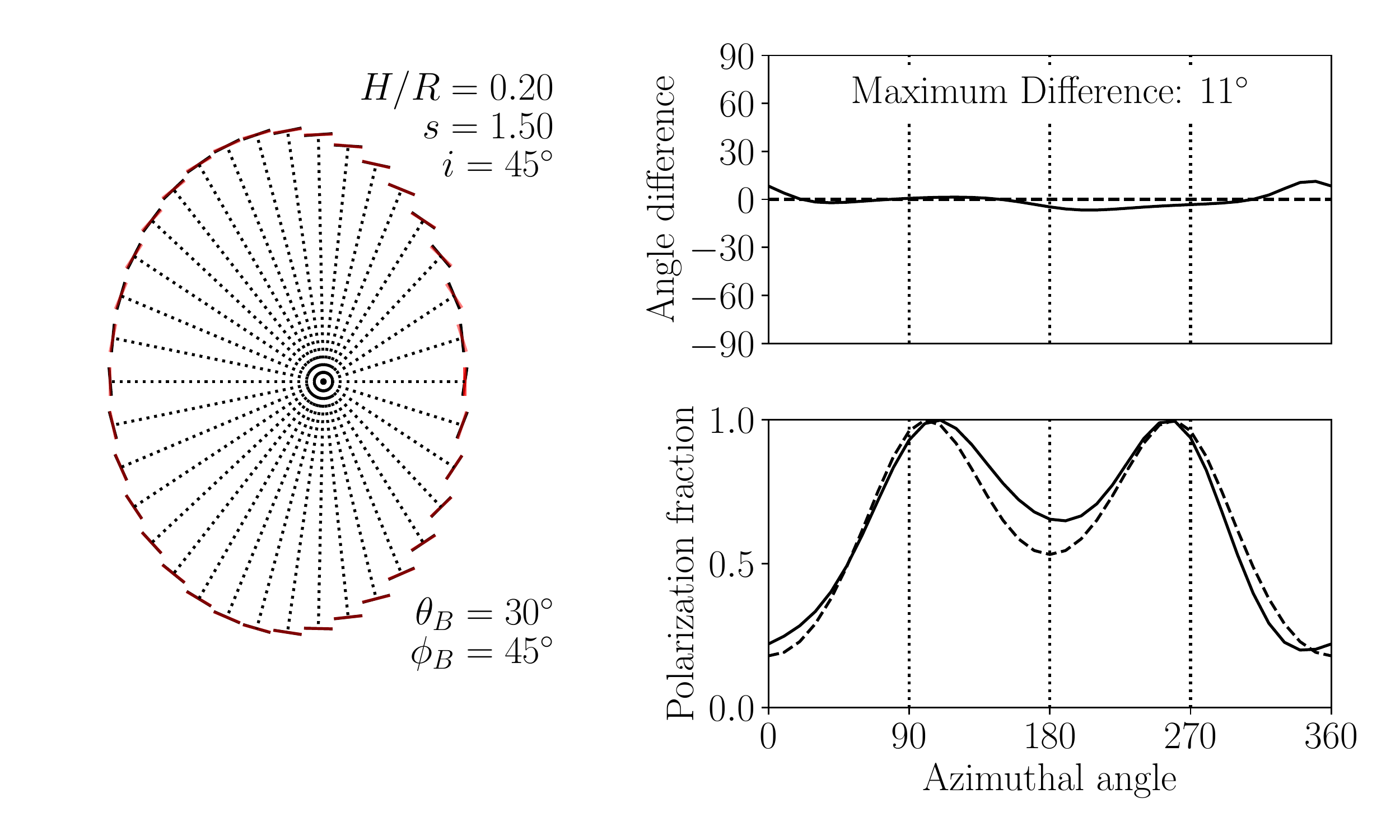}
    \caption{The same as Figure~\ref{fig:disk_fid}, but for Model 2. This represents a less inclined disk model with a moderate magnetic field configuration. There is no polarization reversal. The maximum angle difference is only $11^\circ$ but still detectable at $3\sigma$ level if SNR$>7.8$ for the Stokes parameters.}
    \label{fig:disk_model2}
\end{figure*}

As an example of smaller inclination angle, we assume $H/R=0.2$, $s=1.5$ and $i=45^\circ$. We also assume a less extreme configuration for the  magnetic field with $\theta_B=30^\circ$ and $\phi_B=45^\circ$. The results for this model (Model 2 hereafter) are shown in Figure~\ref{fig:disk_model2}. We can see that for this less inclined disk model, the maximum angle difference is $11^\circ$. 
Note that the toroidal to poloidal magnetic field ratio is $B_\mathrm{tor}/B_\mathrm{pol}=1/\sqrt{7}\approx 0.38$.
In Section~\ref{ssec:error}, we derive a rough error estimate formula for the angle deviation as $\delta \eta \approx 1/(2\mathrm{SNR})$, with SNR being the signal to noise ratio.
The angle difference of $11^\circ$ can be detected at a signal to noise level of $\delta\eta/\Delta\eta = 0.38\,\mathrm{SNR}$. If we ask for 3$\sigma$ detection for the angle difference, we need only an SNR of $7.8$, easily achievable with VLT/SPHERE. 

\begin{figure}
    \centering
    \includegraphics[width=0.5\textwidth]{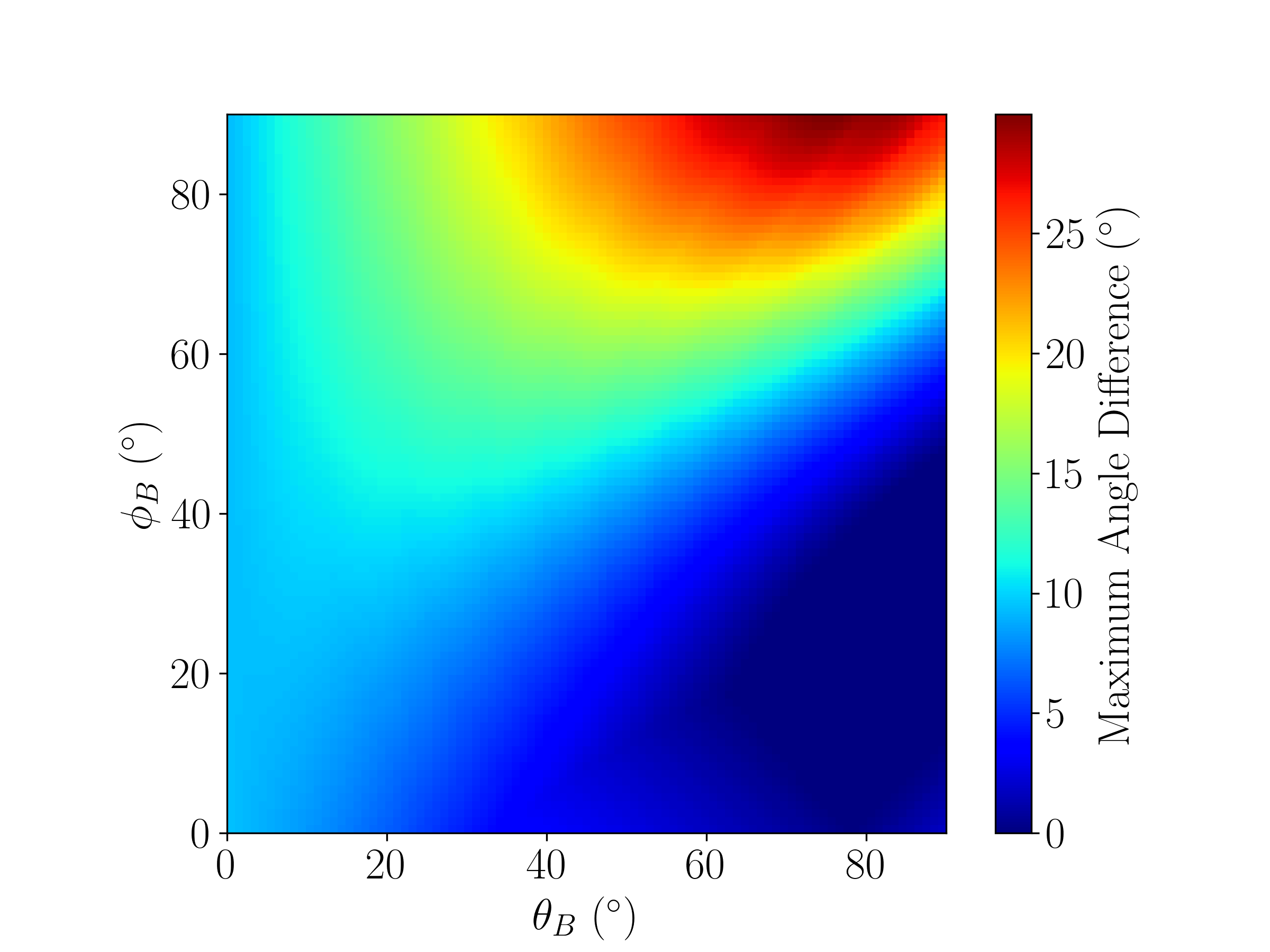}
    \caption{The same as Figure~\ref{fig:Bphase} but for $i=45^\circ$, $H/R=0.2$, $s=1.5$, and without the constant angle difference $\Delta\eta=70^\circ$ contours. Note the difference in the color map.}
    \label{fig:Bphase_model2}
\end{figure}

If we allow the magnetic field configuration to change while fixing the other parameters, we get the maximum angle difference in the $(\theta_B,\phi_B)$ map shown in Figure~\ref{fig:Bphase_model2}. 
We can see that the trend is similar to our fiducial model, except that there is no polarization reversal in this map.
The maximum angle difference increases as we increase the $\phi_B$. 
Even for this moderate inclination angle of $i=45^\circ$, as $\phi_B$ approaches $90^\circ$, the angle difference can easily reach $20^\circ$ or even $30^\circ$. 
The azimuthal angle of the magnetic field, $\phi_B$, also determines the ratio of the toroidal component to the poloidal component of the magnetic field, which is very important in determining the wind launching mechanisms. 
Disks with large magnetization have rigid magnetic field lines, and tend to launch magnetocentrifugal winds with a small toroidal component. 
Weakly magnetized disk, on the other hand, will have magnetic field lines winded up into mostly toroidal configuration first. The disk wind is then launched due to a vertical gradient of magnetic pressure \citep{BaiYe2016}.

\begin{figure}
    \centering
    \includegraphics[width=0.45\textwidth]{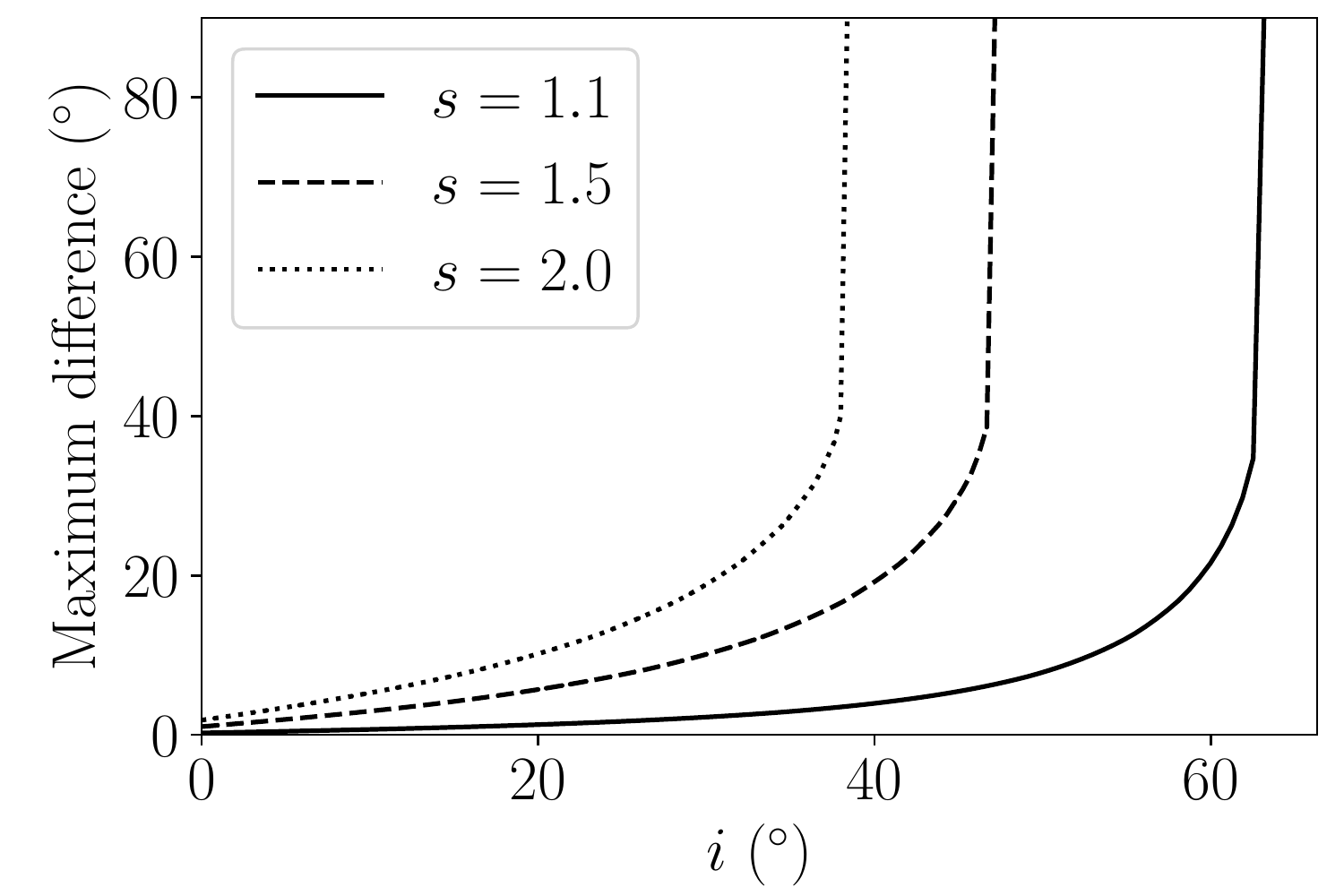}
    \caption{Maximum angle difference as a function of inclination angle $i$, for $H/R=0.2$ and three different dust aspect ratios $s=1.1,\,1.5,$ and $2.0$.}
    \label{fig:approaching}
\end{figure}

The maximum angle differences in the $(\theta_B,\phi_B)$ map can be calculated for different inclination angles $i$ while fixing $H/R=0.2$.
The results for three different dust aspect ratios $s=0.1,\,1.5,$ and $2.0$ are shown in Figure~\ref{fig:approaching}. We can see that the behaviors are similar among different $s$: the maximum angle difference gradually increases before reaching about $30^\circ$, then it  suddenly jumps to $90^\circ$ and enters the polarization reversal regime, the gray region in Figure~\ref{fig:iotac}. So for disks with small inclination angles, say $i<20^\circ$, the deviation from the azimuthal polarization pattern due to grain alignment is likely negligible. 

\section{Analysis in a disk model}
\label{sec:discussion}

\subsection{Disk model adopted}

In this section, we perform a simple analysis of scattering in the disk atmosphere problem. To do so, we adopt a Minimum Mass Solar Nebular model \citep{Weidenschilling1977} with a column density $\Sigma(R) = \Sigma_0 R_\mathrm{au}^{-1.5}$, where $\Sigma_0=10^3\rm\, g/cm^2$ and $R_\mathrm{au}$ is the cylindrical radius in units of au. 
We assume a vertically isothermal temperature profile with $T(R)=T_0\ R_\mathrm{au}^{-0.5}$, with $T_0=300$ K. 
This results in a mildly flared disk with 
$(H_g/R)=0.045\ R_\mathrm{au}^{1/4}$, where $H_g$ is the gas scale height. 
The dust scale height is different from $H_g$ and depends on the grain size, as \citep{Youdin2007}:
\begin{equation}
    H_d(R,a) = H_g(R) \left(1+\frac{\mathrm{St}}{\alpha}\frac{1+2\mathrm{St}}{1+\mathrm{St}}\right)^{-1/2},
    \label{eq:Hd}
\end{equation}
where $\mathrm{St}=\rho_s a/\Sigma$ is the Stokes number that determines how well the dust grains are coupled with the gas, and $\alpha$ is the turbulence parameter, which we take to be $\alpha=10^{-4}$. We assume dust grains have a power-law distribution \citep{MRN}: $N(a)\propto a^{-3.5}$ after vertical integration, between $a_\mathrm{min}=0.01\micron$ and $a_\mathrm{max}=1$ mm. The total column density of the dust grains is $0.01\Sigma$, where a gas-to-dust ratio of $100$ is assumed. 
In practice, we use $100$ bins of dust grains distributed evenly in logarithmic space, each represented by the center of the bin. For each bin, the dust grains follows vertical Gaussian distributions according to a scale height from Equation~\eqref{eq:Hd}.

\subsection{Grain size at $\tau=1$ surface}

The $\tau=1$ surface is the surface where radial optical depth $\tau$ reaches $1$. This is where the stellar light is scattered by the dust grains, and the properties of dust grains at this surface is very important. 

\begin{figure}
    \centering
    \includegraphics[width=0.48\textwidth]{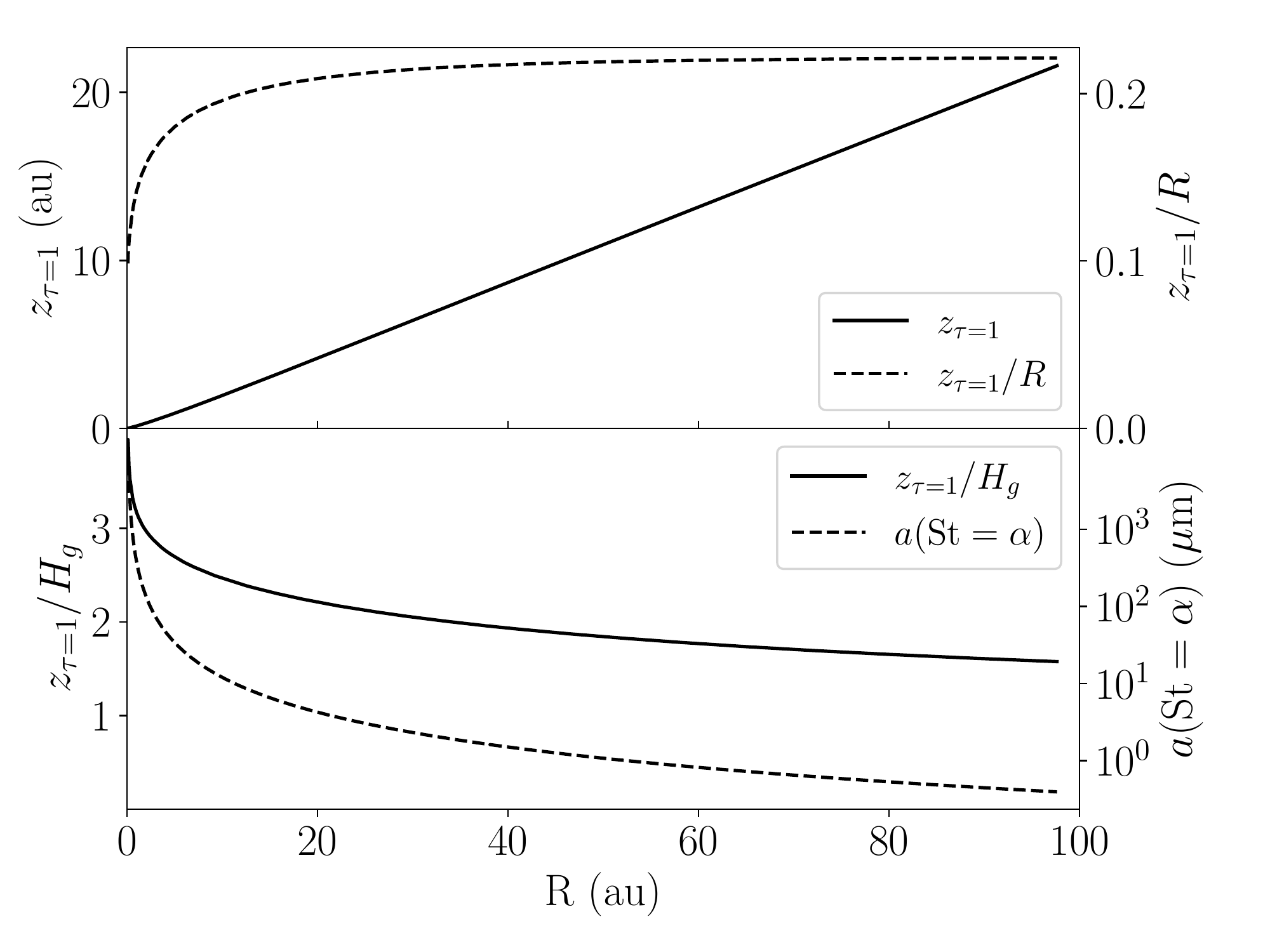}
    \caption{The $\tau=1$ surface. \textit{Up}: The solid line represents the height of $\tau=1$ surface as a function of radius, whereas the dashed line shows the $H/R$ of this surface. \textit{Bottom}: The solid line shows the height of $\tau=1$ surface in terms of local gas scale height. The dashed line represents the critical grain size as a function of radius defined by setting the Stokes number equal to the  viscous parameter $\alpha$.}
    \label{fig:tau1}
\end{figure}

In order to calculate the $\tau=1$ surface, 
we calculate the extinction cross section at $1.5\,\micron$ for grains of different sizes using Mie theory \citep{BH83} through the \texttt{miepython} module\footnote{Credit: Scott Prahl.\\ Available at \url{https://github.com/scottprahl/miepython}}. We then
integrate the optical depth at $\lambda=1.5\,\micron$ radially outward from the center of the disk. The $\tau=1$ contour 
is shown in the upper panel of Figure~\ref{fig:tau1}. 
Also shown in the upper panel as a dashed line is the $H/R$ of the surface.
We can see that the $\tau=1$ surface is largely flat, despite the fact that our disk
model is mildly flared. 
There are two reasons for this flat surface. On the one hand, the dust settles towards the midplane more at large radii due to the less turbulent stirring from the more diffuse gas. This effect can also be viewed in the lower panel of Figure~\ref{fig:tau1}, where the $z_{\tau=1}/H_g$ is plotted against $R$ as a solid line. 
We can see that the $\tau=1$ surface in terms of the gas scale height $H_g$ gradually decreases.
On the other hand, and probably more importantly, the outer regions are blocked or ``shadowed'' by the inner regions, so that the $\tau=1$ surface can never bend towards midplane. 

Note that \cite{Avenhaus2018} finds that the protoplanetary disks are moderately flared, as opposed to being flat in our model. 
In our turbulent stirring model, if we fix the grain size, the Stokes number goes as $\mathrm{St}\sim R^{\gamma}$, where $-\gamma=-1.5$ is the power-law index for the adopted column density profile. 
In the limit of $\alpha\ll \mathrm{St}\ll 1$, we have $H_d/H_g\sim \mathrm{St}^{-1/2}\sim R^{-\gamma/2}$. 
The gas scale height is $H_g/R\sim R^{(1-q)/2}$, with $-q=-0.5$ being the power-law
index for the temperature profile. So we have, for fixing grain size $a$, $H_d/R\sim R^{(1/2)(1-q-\gamma)}$. 
Since we have adopted $q=0.5$, $\gamma=1.5$, the $H_d/R$ decreases with $R$, and the dust at larger radii is shadowed by the dust at inner
radii.
If we require the near-IR scattering surface to be flared as well, we need $q+\gamma<1$. This is very hard to achieve. 
The requirement may be alleviated with the introduction of radial variation of turbulent $\alpha$. If $\alpha$ goes as 
$R^{\omega}$, i.e. the disk is more turbulent at a larger radius, the above constraint becomes $q+\gamma<1+\omega$. 
Another potential and more likely way to make a flaring scattering surface is to abandon the turbulent stirring model and introduce a
disk wind, which may entrain small grains and make the disk appear flaring.
We will not discuss these alternatives in more detail in what follows. The turbulent stirring model we introduce here is 
only for illustrative proposes and to lay the foundations for the following discussions on synthetic maps and grain 
alignments, both of which are not sensitive to whether the disk is flared or not.

In the lower panel of Figure~\ref{fig:tau1}, we also plot the grain size with $\mathrm{St}=\alpha$ as a function of $R$. It characterizes the maximum size of the grains that can be stirred up to a height comparable to the gas scale height (c.f. Equation~\ref{eq:Hd}). 

\begin{figure}
    \centering
    \includegraphics[width=0.48\textwidth]{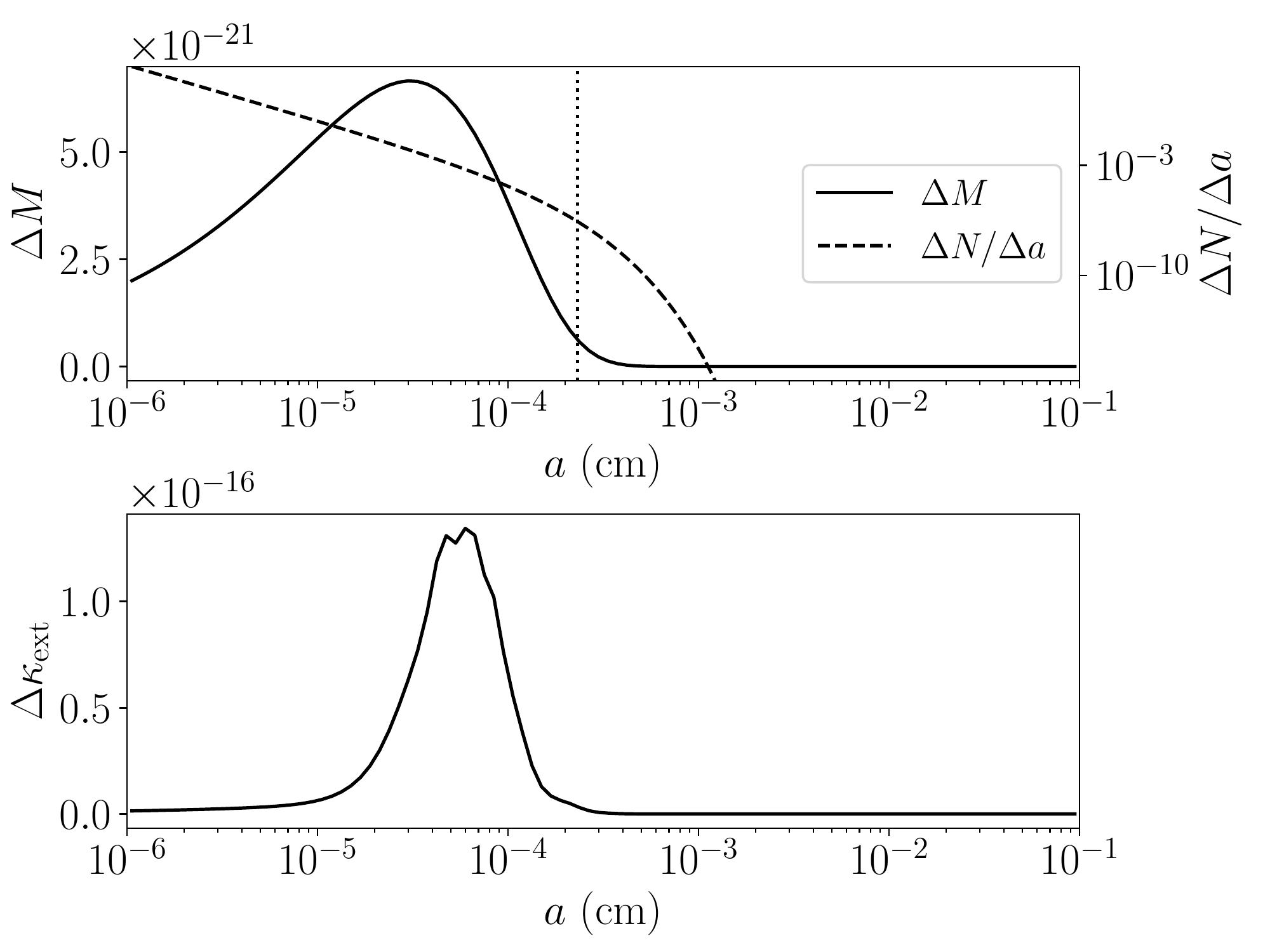}
    \caption{The mass, number density and extinction opacity as a function of different grain sizes at the $\tau=1$ surface of $R=30$ au.}
    \label{fig:R30}
\end{figure}

To view the dust distribution at $\tau=1$ surface more clearly, we plot the mass, number density, and extinction cross section at $R=30$ au for each grain size bin in Figure~\ref{fig:R30}.
The mass $\Delta M$ is the mass density of dust grains between $a$ and $a+\Delta a$. The number density is $\Delta N/\Delta a=\Delta M/(m_a \Delta a)$, with $m_a=\rho_s(4\pi/3)a^3$ being the mass of a grain with radius $a$. The extinction cross section is defined as $\Delta \sigma_\mathrm{ext}=\Delta M\kappa_\mathrm{ext}(a)$, with $\kappa_\mathrm{ext}(a)$ being the extinction opacity for the grains of radius $a$.

At $R=30$ au, we have $z_{\tau=1}= 6.4$ au, and $a(\mathrm{St}=\alpha)=2.3\micron$. When the grain size increases beyond $1\micron$, the mass and the number density drop very quickly. The extinction is also dominated by grains with radius of $\sim 0.6\micron$, or size parameter of $2\pi a/\lambda\sim 2.5$. This justifies our discussion on grain size in Section~\ref{ssec:large} and the adoption of dipole approximation in most of this work.

\subsection{Synthetic maps}
\begin{figure*}
    \centering
    \includegraphics[width=0.95\textwidth]{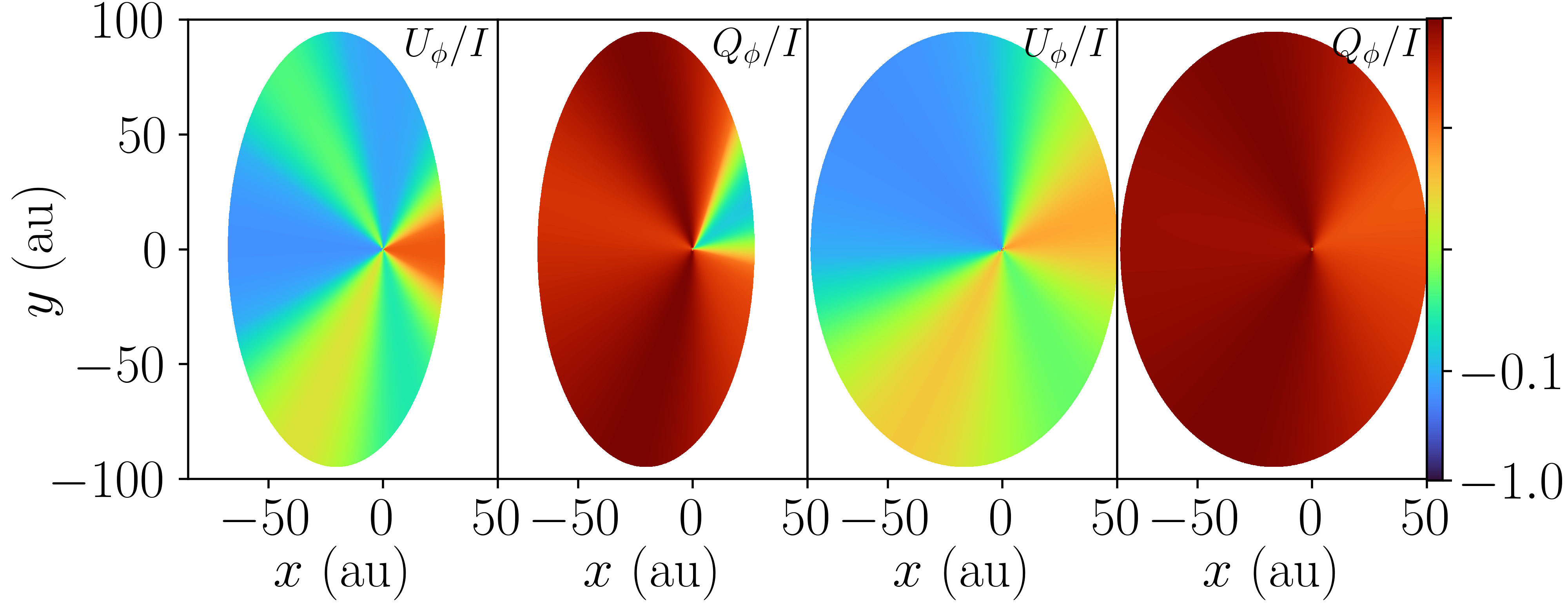}
    \caption{Synthetic maps. The left two panels are $U_\phi/I$ and $Q_\phi/I$ for Model 1 with $i=60^\circ$. The right two panels are the same but for Model 2 with $i=45^\circ$. Note that the colormap is the same for all panels and uses symmetric logarithmic normalization while being linear between $-0.1$ and $0.1$.}
    \label{fig:qumap}
\end{figure*}

With the $\tau=1$ surface and the corresponding $H/R$ obtained, we can calculate the Stokes parameters at each location and generate synthetic maps. In order to focus on the deviation from the azimuthal pattern, we choose to show maps of the ratios of azimuthal Stokes parameters, $U_\phi/I$ and $Q_\phi/I$. The results for Model 1 and Model 2 are shown in Figure~\ref{fig:qumap}. In order to make small values more visible while using the same colormap for all panels, we adopt symmetric logarithmic normalization with a linear scale between $-0.1$ and $0.1$.

For Model 1, we have polarization reversal (polarization in radial direction), which manifests itself as a negative wedge in the $Q_\phi/I$ map. The magnitude of the negative wedge is small ($\sim 0.05$) because the polarization reversal coincide with low polarization points. The $U_\phi/I$ in Model 1 can be as large as $\sim 0.15$, and can be both positive and negative. 

For Model 2, we do not have polarization reversal so that all $Q_\phi/I$ are positive. The magnitude of $U_\phi/I$ is also smaller compared to the more inclined Model 1 and stays below $\sim 0.05$ over most of the disk.

Last but not the least, the sign of $U_\phi$ depends on the direction of the toroidal magnetic field. If we change the $\phi_B$ to $-\phi_B$, the $Q_\phi/I$ maps are unaffected but the $U_\phi/I$ maps will change signs everywhere while keeping the magnitude the same. This behaviour, alongside information on disk rotation directions, can be used to distinguish it from other mechanisms that produce $U_\phi$, such as through multiple scattering \citep{Canovas2015}. 
We will discuss how to distinguish our $U_\phi$-producing mechanism from others in more detail in Section~\ref{ssec:distinguish}.

\subsection{Grain alignment at $\tau=1$ surface}

In this section, we briefly discuss the grain alignment at the $\tau=1$ surface based on timescales of several most relavent processes. The discussion is similar to \cite{Tazaki2017} and \cite{Yang2021}. 

\begin{figure}
    \centering
    \includegraphics[width=0.45\textwidth]{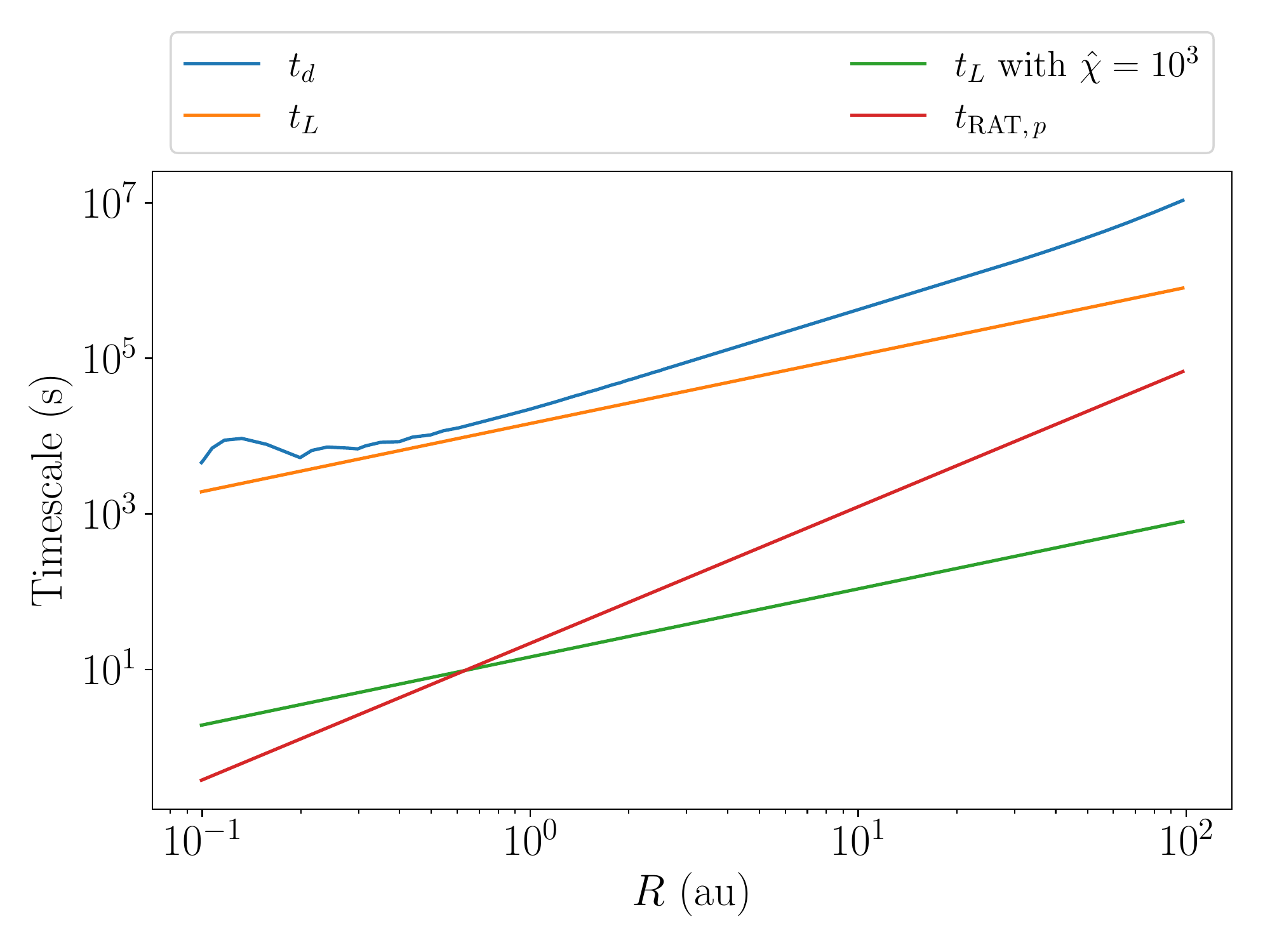}
    \caption{Timescale comparisons for grain alignment}
    \label{fig:timescales}
\end{figure}

1. The gaseous damping timescale: it determines how fast random collisions with gas particles disalign dust grains and is given by:
\begin{equation}
    \begin{split}
    t_\mathrm{d} = & 7.1\times 10^{5}~\mathrm{s}\times \left(\frac{\rho_s}{3~\mathrm{g/cm^3}}\right)\\
    &\times \left(\frac{a}{1~\mathrm{\mu m}}\right)\left(\frac{n_{\rm g}}{10^9~\mathrm{cm^{-3}}}\right)^{-1}\left(\frac{T_g}{85~\mathrm{K}}\right)^{-1/2},
    \end{split}
    \label{eq:td}
\end{equation}
where $\rho_s$ is the solid density of dust grains, $n_g$ is the number density of gas molecules (assuming mean molecular weight of $2.3$).
For our adopted disk model at $\tau=1$ surface and $1\rm\, \mu m$ dust grains, the gaseous damping timescale is plotted as a blue curve in Figure~\ref{fig:timescales}. 

2. The Larmor precession timescale: it is the timescale that rotating dust grains with magnetic moment due to the Barnett effect \citep{Barnett1915} precess around an external magnetic field:
\begin{equation}
    \begin{split}
    t_L = 1.5\times 10^6\,\mathrm{s} 
    &\times \hat{\chi}^{-1} \hat{\rho}_s
    \left( \frac{T_d}{85~\mathrm{K}} \right) \\
    &\left( \frac{B}{5~\mathrm{mG}} \right)^{-1} \left( \frac{a}{1~\mathrm{\mu m}} \right)^2,
    \end{split}
    \label{eq:tL}
\end{equation}
where $T_d$ is the dust temperature which we take to be the same as the midplane gas temperature prescribed above (vertically isothermal).
For the magnetic field, following \citep{Yang2021}, we adopt the estimate from \citep{Bai2011}:
\begin{equation}
B=1.0\,\mathrm{G}\times \left(\frac{\dot{M}}{10^{-8}\rm\,M_\sun/yr}\right)^{1/2}\left(\frac{r}{1\rm\,AU}\right)^{-11/8},
\label{eq:Bmod}
\end{equation}
where we have assumed the mass accretion rate $\dot{M}=10^{-8}\ M_\odot\,\mathrm{yr}^{-1}$, typical for classical T Tauri stars \citep{Hartmann2016}.

The Larmor precession timescale is plotted as an orange curve in Figure~\ref{fig:timescales}. It is possible for dust grains to carry superparamagnetic inclusions \citep{JS1967}, which can shorten the Larmor precession timescale by up to a factor of $\hat{\chi}\sim10^3$ \citep{Yang2021}. We also plot the Larmor precession timescale with superparamagnetic inclusions with $\hat{\chi}=10^3$ as a green curve.

3. RAT precession timescale: it is the timescale for the precession due to the Radiative Alignment Torque \citep{LH2007}. It can be estimated as \citep{Tazaki2017}:
\begin{equation}
\begin{split}
t_{\rm rad,\,p} &= 8.8 \times 10^4\,\mathrm{s}\times \hat{\rho}_s^{1/2} \hat{s}^{-1/3} \left(\frac{a}{1\rm\, \mu m}\right)^{1/2} \left(\frac{T_d}{85\rm\, K}\right)^{1/2}\\
&\left(\frac{u_\mathrm{rad}}{10^6~u_\mathrm{ISRF}}\right)^{-1} \left(\frac{\bar{\lambda}}{0.89\rm\, \mu m}\right)^{-1}
\left(\frac{\gamma \overline{|Q_\Gamma|}}{0.4}\right)^{-1},
\end{split}
\label{eq:trad}
\end{equation}
where $\bar{\lambda}$ is the energy weighted averaged wavelength of the radiation, $\gamma$ is the anisotropy of the radiation, and
$u_\mathrm{ISRF}=8.64\times10^{-13}$ is the interstellar radiation energy density \citep{Mathis1983}. 
We consider only the stellar light, hence $\gamma=1$.

For radiation energy density $u_\mathrm{rad}$, we assume solar parameters, with effective temperature of $\sim 6000$ K and bolometric luminosity $L=L_\odot$. This yields:
\begin{equation}
u_\mathrm{rad}=\frac{L}{4\pi R^2 c} 
= 4.564\times10^{-5} \, R_\mathrm{au}^{-2} \, \mathrm{erg\,cm^{-3}}\,\,,
\end{equation}
and $\bar{\lambda}=0.89\micron$. Since $\bar{\lambda}\leq 1.8 a$, we have $|Q_\Gamma|\approx 0.4$ \citep{LH2007}. The RAT precession timescale is plotted as a red curve in Figure~\ref{fig:timescales}.

From Figure~\ref{fig:timescales}, we can see that $t_{\mathrm{RAT},p}$ is always smaller than $t_d$, indicating efficient Radiative Alignment torque. If the dust grains are of regular paramagnetic materials (orange curve), we also have $t_L>t_{\mathrm{RAT},p}$. In this case, we have $k$-RAT, i.e. grains aligned with radiation flux. As a result, the dust grains are aligned with short axes along the stellar light direction. 
As a result, the dust grains would look round from the star and they scatter stellar light exactly the same as spherical dust grains in the dipole regime with small particles. 
Hence we expect no deviation from the azimuthal pattern if dust grains are aligned with $k$-RAT.

If the dust grains possess superparamagnetic inclusions (SPIs), the Larmor precession timescales can be reduced by a factor up to about $10^3$ \citep{Yang2021}. In this case (the green curve), we have $t_L>t_{\mathrm{RAT},p}$ outside a radius of $0.7$ au. 
So it is likely that superparamagnetic dust grains are aligned with the  magnetic field rather than the radiation flux for the majority of the disk at tens of au scale. 
As SPI-candidates are seen in meteorites \citep{Goodman1995}, it is possible that dust grains in the disk atmosphere possess SPIs as well. The near-IR wavelength scattering polarimetry of protoplanetary disks can be an excellent probe for the existence of SPIs, which will help understanding magnetic alignment of large dust grains in other environments.

Lastly, it is worth mentioning that the internal relaxation can also be problematic in disk atmosphere. \cite{Tazaki2017} estimated the internal relaxation timescale as a function of grain radius in their Figure~2 and showed that the internal relaxation timescale of $1\rm\, \mu m$ dust grains can be tens of years, longer than any of the timescales considered in Figure~\ref{fig:timescales}. In this case, the degree of alignment may be reduced due to the lack of internal relaxation \citep{Hoang2009,Tazaki2017}. More detailed and quantitative discussion and modeling on grain alignment is beyond the scope of this paper and will be deferred to future investigations.

\section{Discussion}
\label{sec:added_discussion}

\subsection{Detectability}
\label{ssec:error}
In previous sections, we have shown that the angle deviation from azimuthal polarization can easily reach $10^\circ$ or more for a moderately inclined disk. Here we discuss the detectability of such angle deviations.

The angle with azimuthal direction can be calculated through $\eta = (1/2)\mathrm{arctan2}(U_\phi,Q_\phi)$. Without loss of generality, we limit our discussion here to the quadrant where $Q_\phi>0, U_\phi>0$, 
so that $\eta = (1/2)\mathrm{arctan}(U_\phi/Q_\phi)$.  The total differential is then:
\begin{equation}
\delta \eta = \frac{Q_\phi\delta U_\phi-U_\phi\delta Q_\phi}{2(Q_\phi^2+U_\phi^2)}.
\end{equation}
We can see that the above expression is on the order of $1/(2\mathrm{SNR})$, where SNR is the signal to noise ratio of polarized intensity defined as $\delta \mathrm{PI}/\mathrm{PI}$, with PI being the polarized intensity. 
This estimate can be made more accurate if we focus on the deviation from the azimuthal pattern where $Q_\phi=\mathrm{PI}, U_\phi=0$, so that $\delta \eta = \delta U_\phi / 2Q_\phi = 1/(2\mathrm{SNR})$. 

For the $10^\circ$ angle deviation we obtained before, we need SNR$\le 7.8$. In the survey presented by \cite{Avenhaus2018}, the SNR is better than $20$ for most cases, which translates into an error in angle of $\delta\eta\sim 1.4^\circ$.
So if the dust is aligned with the magnetic field in the atmosphere of a moderately inclined disk, we should be able to detect the deviation from the azimuthal polarization pattern as predicted by this work.

\subsection{Distinguishing different mechanisms}
\label{ssec:distinguish}
The most important feature of our polarization mechanism is that we relies on elongated dust grains that are aligned with magnetic fields. 
If a mechanism that produces near-IR scattering polarization accounts for only spherical dust grains or non-spherical dust grains but without grain alignment, the optical properties of the ensemble of dust grains will have spherical symmetry, i.e. the scattering matrix is solely a function of scattering angle. 
Under this assumption, for light last scattered at the near-side (the right point in Figure~\ref{fig:disk_fid}) or at the far-side (the left point in Figure~\ref{fig:disk_fid}), the geometry of the scattering problem is symmetric between 
up and down\footnote{We have also assumed axis-symmetric geometry for the PPD. If there is any structures in the PPD, the local radiation field will not have symmetry between up and down. Mechanisms like multiple scatterings \cite{Canovas2015} that
relies on the anisotropy of the local radiation field will also produce non-zero $U_\phi$ component. We will ignore such cases for now.}. 
As a result, the end polarization orientation would either be along the radial direction or along the azimuthal direction with no other possible outcomes, which means $U_\phi=0$.
In contrast, the $U_\phi$ is maximized near the near-side and the far-side points in our models (c.f. Figure~\ref{fig:qumap}). 

\begin{figure}
    \centering
    \includegraphics[width=0.45\textwidth]{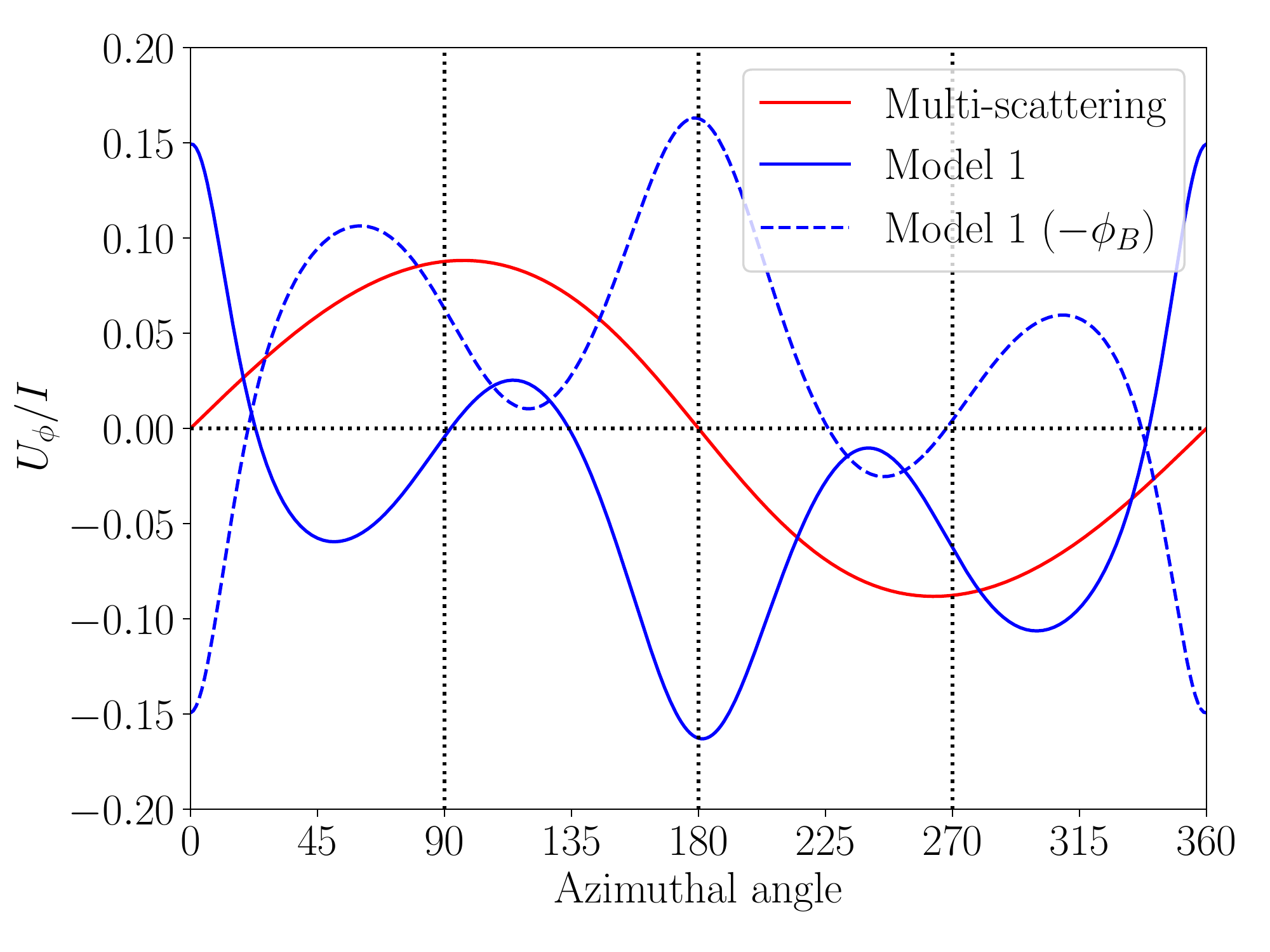}
    \caption{The $U_\phi/I$ profiles. The red curve shows the $U_\phi/I$ for multiple scattering, 
    using a simple double-scattering approximation, assuming an inclination angle of $60^\circ$. 
    See text for discussions. The blue solid curve comes from the fiducial model, Model 1, with $i=60^\circ$. The blue dashed curve is similar to Model 1, but with $\phi_B=-90^\circ$. 
    This changes the helicity of the assumed magnetic field configurations, but nothing else, 
    resulting in $U_\phi$ flips its sign in the whole disk. This is a unique feature for polarization
    produced by magnetic alignments. }
    \label{fig:ms_comp}
\end{figure}

The main alternative discussed in the literature so far is the multiple scattering at high optical depth and high inclination \citep{Canovas2015}. 
Here we give a heuristic argument on how this mechanism works.
If we take single scattering of spherical dust grains as the zeroth order problem, the first order problem will be the photons scattered twice before they reach our telescope. 
As discussed above, the zeroth order problem considering only single scattering cannot produce $U_\phi$. The $U_\phi$ is then produced primarily by the first order problem with double scattering.
Since the disk atmosphere is optically thick at near-IR, the first scattering site cannot
be too far from the second scattering site. We shall refer to the particle at first scattering site as particle A and the particle at the second scattering site as particle B. 
The light coming from particle B is then what we observe. 
In this first order double scattering problem, the local anisotropy of radiation field as viewed at the location of particle B is what determines the polarization state
of the scattered light. If we further ignores the polarization of the light between particle A and B and treat the light scattered by the particle A as non-polarized, the problem reduces to a problem that is the same as the self-scattering
problem at high optical depth at (sub)millemeter wavelengths. The particle A is comparable with the original source of the dust thermal emission at (sub)millemeter wavelengths, and the particle B is the scattering particle of the self-scattering problem.
As discussed by \cite{Yang2017}, the polarization orientation is along the ``minor axis'' of the local disk surface, the direction that is coplanar with both the final scattering direction and 
normal direction of the local disk surface (see Figure 1 of \citealt{Yang2017} for a schematic illustration and the related texts). 
With this simple model, we can calculate the $U_\phi/I$ as a function of the azimuthal angle in the disk frame. The results assuming $i=60^\circ$ are shown as the red curves in Figure~\ref{fig:ms_comp}. 
Note that the absolute values of $U_\phi/I$ from the above simple double scattering model is arbitrary in the sense that the contribution from single scattering is not taken into account in this simple double scattering model. 
The single scattering does not produce $U_\phi$, so it should not affect the overall profile, if azimuthal variation in single scattering is ignored. Higher order scattering events may have more significant contributions which are not taken into account here.
We choose to multiply the whole curve by $0.2$ to make the results comparable with those from the magnetic alignment.
Despite the simplicity of this model, it still captures most of the physics and the result agrees with \cite{Canovas2015}'s moderate opacity model very well. 
The most important features of $U_\phi$ generated by multiple scattering is that the $U_\phi$ is opposite between symmetric points with respect to the disk minor axis ($0^\circ-180^\circ$ vs. $180^\circ-360^\circ$; c.f. Figure 2 of \citealt{Canovas2015}). 

For comparison, we also plot the the $U_\phi$ for Model 1 with $i=60^\circ$ as blue curves. We can see that the $U_\phi$ is maximized at the near and far sides, with near side having positive $0.15I$ and the far side having 
$-0.15I$. If we change the helicity of the magnetic field by changing $\phi_B$ to $-\phi_B$, the $U_\phi$ at the near and far sides will also flip to their opposite values. The curve for $-\phi_B$ is plotted as blue dashed 
curve in Figure~\ref{fig:ms_comp}. We can see that the solid and dashed blue curves are symmetric with respect to the middle point. The $U_\phi$ maps presented in Figure~\ref{fig:qumap} can be changed to the magnetic field 
configuration with opposite helicity by multiplying the whole map by $-1$ and then flip upside down. This dependence on the helicity of magnetic fields can be very important in distinguishing our mechanisms from others.
If we can infer the helicity of the magnetic fields through the rotation curves in the outflows or jets, we can then check against near-IR scattered polarimetry and see if there is any magnetic field signatures
and see if the predicted $U_\phi$ map for given helicity of magnetic fields agrees with observations.

\subsection{Potential sources with aligned grains}

Before going into specific systems, we summarize our discussions on the difference between our mechanism that relies on elongated dust grains aligned with magnetic fields and the other mechanisms that rely only on scattering
by spherical dust grains. For axis-symmetric system, alternative mechanisms with only spherical dust grains will have $U_\phi$ map being \textit{symmetric with respect to the minor axis of the disk.} The symmetry is to be 
considered in the sense of opposite signs. That is to say, if there is a structure of positive $U_\phi$ in one side of the disk, there has to exist the exact same structure of negative $U_\phi$ in the other side of the disk,
with these two structures being mirror symmetric with respect to the minor axis. For our mechanism that relies on elongated dust grains, there is no such requirement, and the $U_\phi$ map can even be dominated by either positive or negative $U_\phi$ (c.f. Figure~\ref{fig:qumap}). 
\textit{The asymmetry in the $U_\phi$ map} with respect to minor axis and/or \textit{predominant positive/negative $U_\phi$ in axis-symmetric systems} are both signs of our mechanism. 

With this in mind, we find that CU Cha, HD 169142, MWC 614, Hen 3-365, and HD 142527 are some good candidates in the 
Gemini-LIGHTS survey \citep{Rich2022}.
They all show clear deviations from mirror symmetry expected for scattering by only spherical dust grains.
In addition, the outer disk of HD 142527 are predominantly positive. 
Similarly HD 169142 is also predominantly positive. 
HD 34700 A has $U_\phi$ maximized along the minor axis, which cannot be explained by spherical dust grains.
MWC 614 has clear asymmetry in $U_\phi$ and slightly more positive than negative $U_\phi$. 
We include Hen 3-365 (HD 87643) as a good candidate, despite its non-axisymmetric structures which complicate the 
interpretation. 
In addition to the substantial asymmetry in the $U_\phi$ image, Hen 3-365 has a wedge of negative $Q_\phi$ (see also \citealt{Laws2020}), similar to the 
Model 1 (c.f. Figure~\ref{fig:qumap}).  
In the DARTTS-S survey \citep{Avenhaus2018}, we find V4046 Sgr and DoAr 44 as candidates based on their asymmetric $U_\phi$ images. 

We would like to note that the calibration of near-IR polarimetry data is a very complicated process. Part of the 
calibration involves correction for instrumental polarization, stellar polarization, and/or foreground interstellar 
contamination. Since there is no priori knowledge on what the stellar polarization and interstellar polarization are, 
a parameterized approach is usually adopted 
\citep{Avenhaus2018}. How these effects affect the detection of the signals predicted in this work remains to be determined. 

\section{Summary}
\label{sec:summary}

In this paper, we have studied the scattering of the near-IR stellar light by small dust grains, that are aligned with respect to the magnetic fields, in the atmosphere of a protoplanetary disk. We focused on the polarization orientation of the scattered light and showed that the deviation from the often assumed azimuthal pattern can be significant. The main findings are as follows.

\begin{enumerate}

\item We calculated the polarization pattern in a disk frame (DF). We focused on two models: Model 1 with a relatively large inclination angle ($i=60^\circ$) and a large toroidal magnetic field component and Model 2 with moderate parameters ($i=45^\circ$). The Model 1 has polarization reversal, i.e. the scattered light is polarized in the radial rather than azimuthal direction, at certain locations. The Model 2 doesn't have polarization reversal, but still has a maximum angle difference of $11^\circ$, detectable if we have $\mathrm{SNR}>7.8$ for Stokes parameters. 

\item We gave a geometric explanation of the polarization reversal in Section~\ref{ssec:toroidal}. We showed that the polarization reversal is almost inevitable for disks with large inclination angles $i>75^\circ$, regardless of the magnetic field configuration. 

\item The angle difference strongly depends on the field configuration. In particular, it increases with an increasing toroidal component of the magnetic field. Hence it can be used to probe the launching mechanism of magnetized disk wind.

\item With a simple minimum mass solar nebular model, we studied the $\tau=1$ surface for scattering near-IR stellar light and found that the maximum grain size there is on the order of $1\rm\, \mu m$, assuming a turbulent viscosity of $\alpha=10^{-4}$. This justifies the focus of this initial study on relatively small grains. 

\item We calculated synthetic $U_\phi/I$ and $Q_\phi/I$ maps for the two disk models on the $\tau=1$ surface. The peak $U_\phi/I$ is on the order of $\sim 0.15$ and $\sim 0.05$ for Model 1 and 2, respectively. Interestingly, the $U_\phi/I$ is reversed with magnitude unaffected if we change the azimuthal direction of the magnetic field (through $\phi_B\to -\phi_B$). This effect, together with disk rotation information, can be used to distinguish our mechanism from other $U_\phi$-producing mechanisms.

\item We conducted a grain alignment analysis at the $\tau=1$ surface. We found that Radiative Alignment Torque should be operating. For regular paramagnetic dust grains, our model favors $k$-RAT, i.e. grains aligned with radial stellar light. If grains possess substantial superparamagnetic inclusions, $B$-RAT becomes likely. 

\item We compared the azimuthal profiles of $U_\phi$ between our model and an alternative model that relies on multiple scattering of spherical dust grains. We argue that a spatial distribution of $U_\phi$ that is predominantly positive or negative and/or asymmetric with respective to the minor axis of an intrisically axisymmetric disk are are signals of aligned elongated dust grains. We identified a handful of systems in the existing literature that are potential targets to look for magnetically aligned grains in future studies.
\end{enumerate}

\section*{Acknowledgements}

We thank the referee for a detailed and constructive report that helped improving our manuscript significantly.
The authors thank Gregory J. Herczeg and Ruobing Dong for comments and suggestions that helped to improve the manuscript. ZYL is supported in part by NASA 80NSSC20K0533 and NSF AST-1815784.

\bibliographystyle{aasjournal}



\appendix
\counterwithin{figure}{section}
\counterwithin{table}{section}

\section{Angle Difference in Grain Frame}
\label{sec:GF}

In this appendix, we expand on the discussion of the scattering-induced polarization in the grain's frame presented in \S~\ref{sec:basic}.

\begin{figure}
    \centering
    \includegraphics[width=0.4\textwidth]{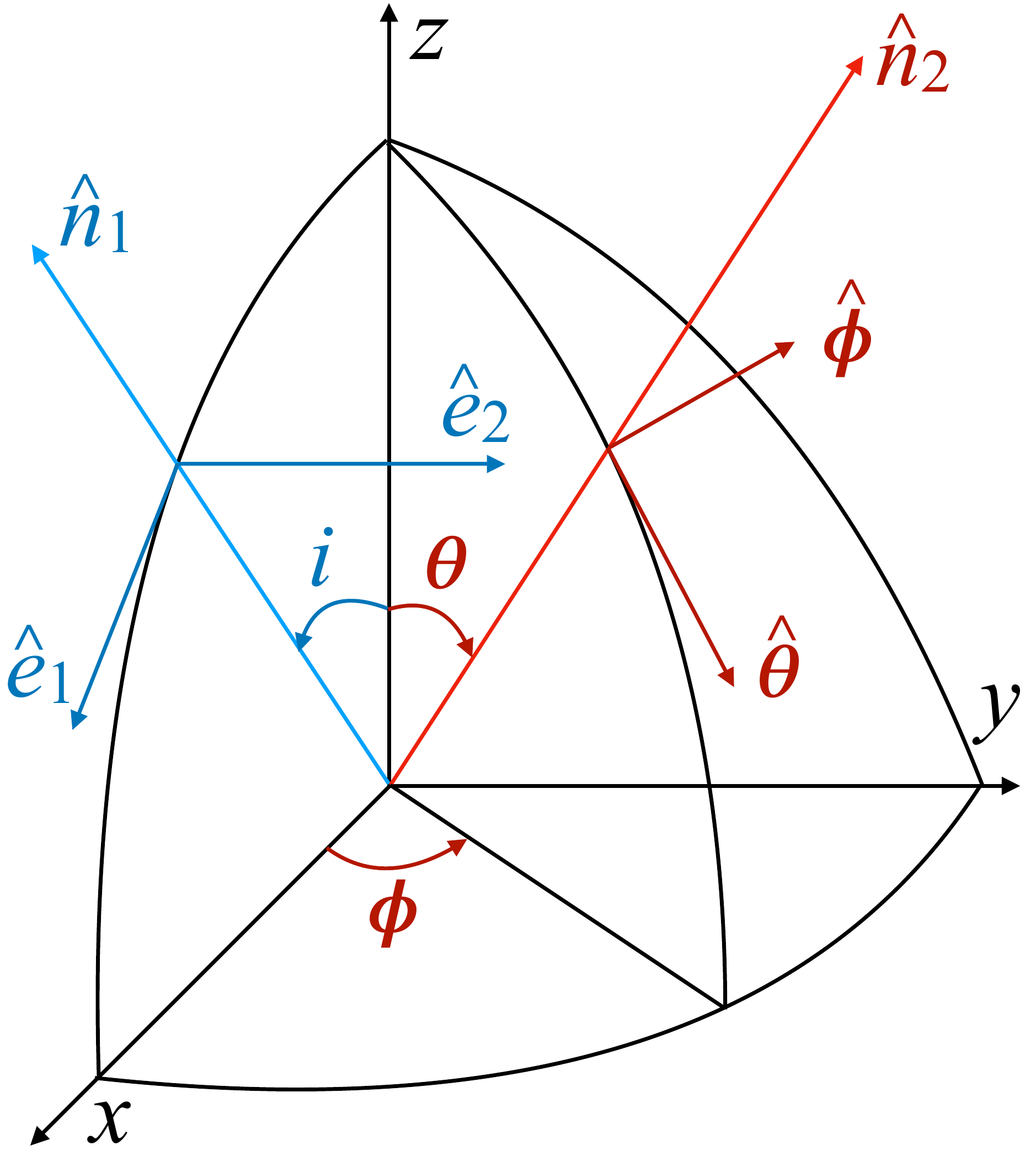}
    \caption{The geometry of our setup in the grain's Frame. The dust grain is located at the origin. The incoming light is propagating along $\hat{n}_1$, making an angle $i$ with $z$ axis in $xz$ plane, with $\hat{e}_1$ and $\hat{e}_2$ as the base vectors for polarization decomposition. The scattered light is propagating along $\hat{n}_2$ with directional angles $(\theta,\phi)$. The base vectors for polarization decomposition are $\hat{\theta}$ and $\hat{\phi}$. See text for more details.}
    \label{fig:GF}
\end{figure}

The geometry of the set-up is shown in Figure~\ref{fig:GF}. The scattering particle is placed at the center of the frame, with $z$ direction along the symmetry axis of the grain. The blue arrow denotes the incoming light $\hat{n}_i$, which is placed in the $xz$ plane, without loss of generality. It makes an angle $i$ with the $z$ axis. The red arrow denotes the scattered light $\hat{n}_s$, defined by two position angles $\theta$ and $\phi$. For the incoming light, its polarization is defined with $\hat{e}_1$ and $\hat{e}_2$. A positive $Q$ corresponds to polarization along $\hat{e}_1$ and a positive $U$ corresponds to polarization along a direction bisecting $\hat{e}_1$ and $\hat{e}_2$. For the scattered light, its polarization is defined with $\hat{\theta}$ and $\hat{\phi}$. Here, a positive $Q$ corresponds to polarization along $\hat{\theta}$ and a positive $U$ corresponds to polarization along a direction bisecting $\hat{\theta}$ and $\hat{\phi}$. The polarization orientation angle $\eta$ is defined in the $\hat{\theta}$-$\hat{\phi}$ plane as the angle starting from $\hat{\theta}$ and increasing towards $\hat{\phi}$ counterclockwise, with values between $0^\circ$ to $180^\circ$. 

In Sections \ref{ssec:spherical}-\ref{ssec:composition}, we will limit our discussions to the dipole approximation (for small grains) to gain a better understanding. The calculation is done with the electrostatic approximation or dipole approximation \citep{BH83}. See Appendix \ref{apsec:dipole} for a brief description of this method. 
The impact of large dust grains is briefly discussed in Section \ref{ssec:large}.

\subsection{Scattering by Small Spherical Particles}
\label{ssec:spherical}

\begin{figure}
    \centering
    \includegraphics[width=0.49\textwidth]{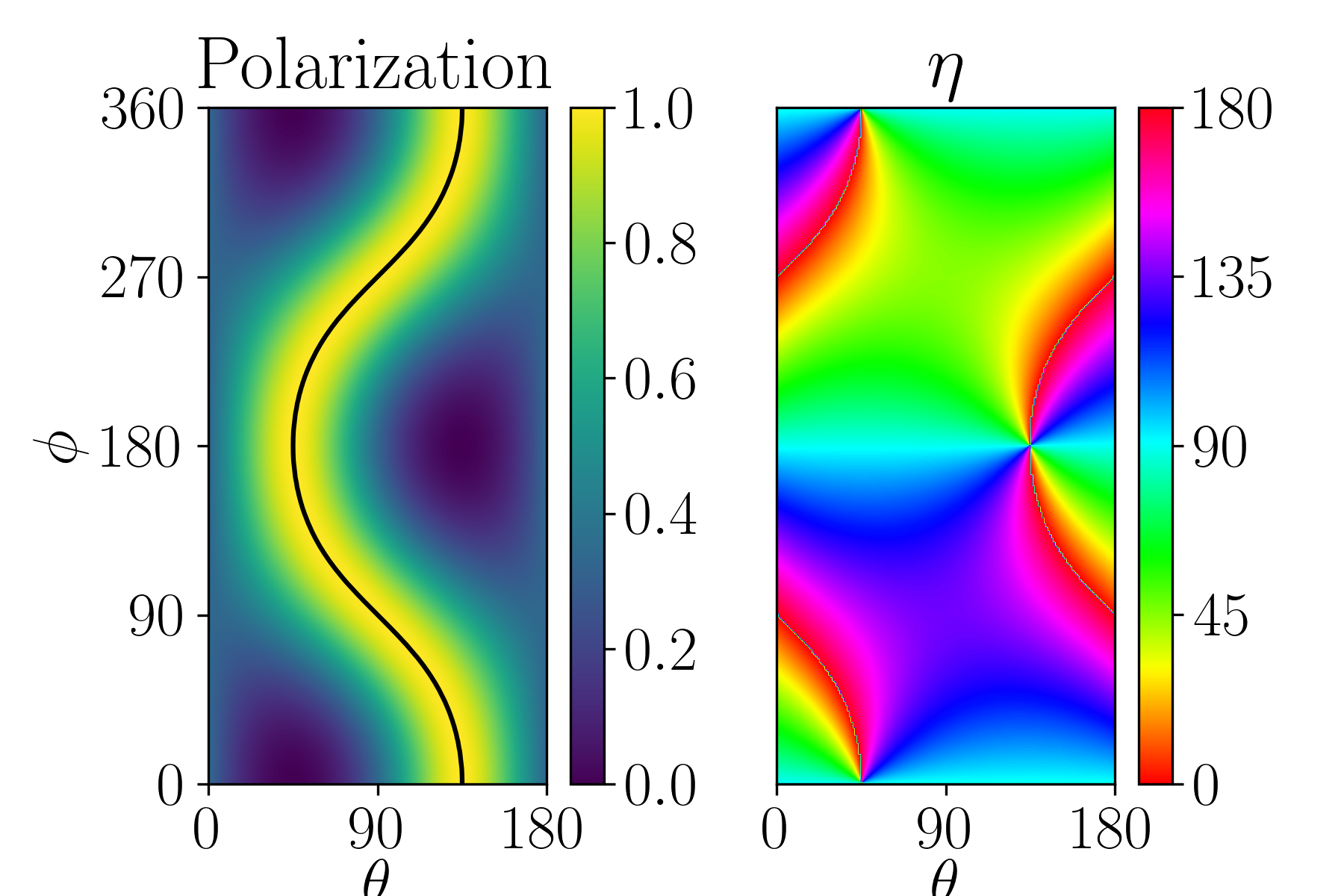}
    \includegraphics[width=0.49\textwidth]{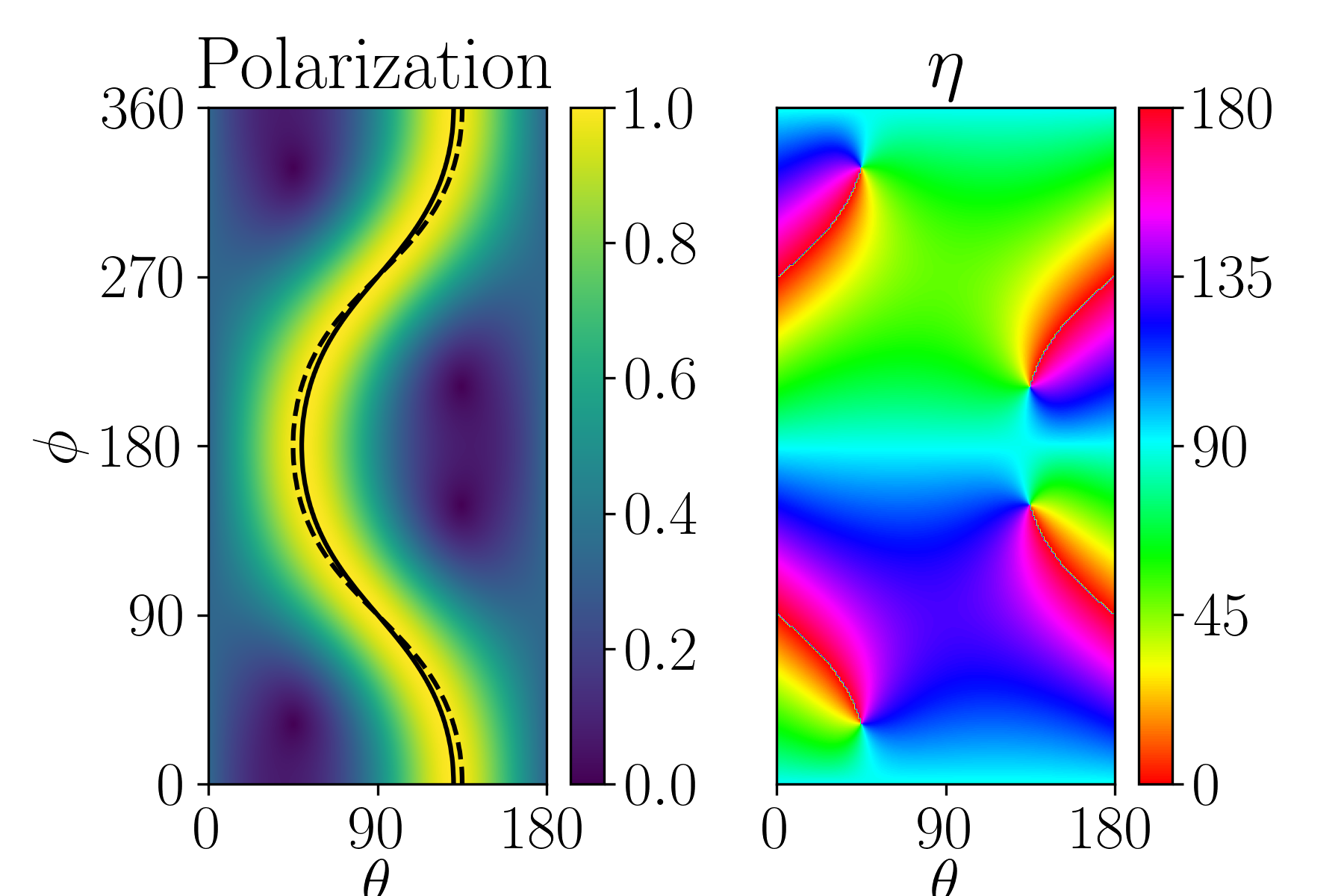}
    \caption{\textit{Left two panels}: The results for small spherical dust grains. The first left panel shows the polarization fraction for different scattered light direction $(\theta,\phi)$. The second left panel shows the polarization orientation angle $\eta$. $\eta=0$ means polarization along $\hat{\theta}$.
    \textit{Right}: The same but for results for our fiducial case: an small oblate dust grain with an aspect ratio $s=1.5$. The scattering inclination angle is $i=45^\circ$.}
    \label{fig:sphere}
\end{figure}

Before calculating the polarization from scattering off elongated dust grains, let's first look at the simpler case for spherical particles when $i=45^\circ$. In the left two panels of Figure~\ref{fig:sphere}, we show the polarization fraction $p$ (first left; defined as $\sqrt{Q^2+U^2}/I$) and the polarization orientation angle $\eta$ (second left) as we change the direction of the scattered light. Note that since the particle is perfectly spherical, the polarization is completely determined by the angle between $\hat{n}_i$ and $\hat{n}_s$ and the results contain no new information but the well known polarization profile $p=\sin^2<\hat{n}_i,\hat{n}_s>/(1+\cos^2<\hat{n}_i,\hat{n}_s>)$ and the fact that the polarization direction is perpendicular to both $\hat{n}_i$ and $\hat{n}_s$. Nonetheless, the spherical case in the left two panels of Figure~\ref{fig:sphere} will serve as a benchmark to help us understand the cases for aspherical dust grains later.

In the first left panel of Figure~\ref{fig:sphere}, we overlay a curve which has $p=1$, i.e. the scattered light is fully polarized. In this small spherical dust grain case, this is the direction that is perpendicular to the incoming light direction.

In the second left panel of Figure~\ref{fig:sphere}, we can clearly see that there exist two singular points, one with $(\theta=45^\circ, \phi=0^\circ)$, and one with $(\theta=135^\circ, \phi=180^\circ)$. These two points correspond to the forward scattering and backward scattering, correspondingly. In these two directions, the polarization fraction in scattered light is $0$, and the polarization direction is ill defined. If we walk around the singular point while fixing $\theta$, the polarization will be along $\hat{\theta}$ direction, in order to be perpendicular to both incoming and scattered light. Similarly, if we walk around the singular point while fixing $\phi$, the polarization will be along $\hat{\phi}$ direction. Note that $\eta=0^\circ$ and $\eta=180^\circ$ correspond to the same polarization orientation.

It is worth mentioning that the polarizability matrix (see Appendix \ref{apsec:dipole} and Equation \ref{eq:polmatrix}) in this spherical case is isotropic and diagonal: $\bar{\alpha}=\mathrm{diag}\{\alpha_s,\alpha_s,\alpha_s\}$, where 
\begin{equation}\alpha_s=a^3\frac{\epsilon - 1}{\epsilon +2},
\end{equation}
with $a$ being the grain size, and $\epsilon$ is the complex dielectric function.

\subsection{Fiducial case}
\label{ssec:fidGF}

Now let's move on the the more interesting case with aspherical dust grains. For our fiducial case, we consider a small dust grain in the dipole regime. In this work, we consider only oblate dust grains characterized by an aspect ratio $s>1$, which we set to $s=1.5$ in our fiducial case. In the grain's frame depicted in Figure~\ref{fig:GF}, the symmetry axis of the dust grain is placed along the $z$ direction. The light makes an angle $i=45^\circ$ with the $z$ direction, the same as the spherical case discussed above. For the fiducial model and most of the models in this paper, we assume the composition from \cite{Birnstiel2018}. It is a mixture of 20\% water ice \citep{Warren2008}, 33\% astronomical silicates \citep{Draine2003}, 7\% troilite \citep{Henning1996}, and 40\% refractory organics \citep{Henning1996} by mass. 
Thoughout this paper, we assume an observing wavelength of $1.5\rm\, \mu m$. The results are largely independent of the specific choice of the wavelength.
The results are shown in the right panels of Figure~\ref{fig:sphere}. 

In the grain's frame, the polarizability matrix is always diagonal as $\mathbf{P}=\mathrm{diag}\{\alpha_1,\alpha_1,\alpha_3\}$, and $|\alpha_1|>|\alpha_3|$ because we assume oblate dust grains. See Appendix~\ref{apsec:dipole} for more details. For an incoming radiation propagating along the $\hat{n}_i$ direction, we can decompose the light to two components: $\mathbf{E}_i=E_1\hat{e}_1+E_2\hat{e}_2$. The dipole excited in response to these two components are:
\begin{equation}
\mathbf{P}_1 = \bar{\alpha}E_1\hat{e}_1 = \alpha_1 E_1 \hat{x}\cos i - \alpha_3 E_1 \hat{z} \sin i,
\end{equation}
and
\begin{equation}
\mathbf{P}_2 = \bar{\alpha}E_2\hat{e}_2 = \alpha_1 E_2\hat{y}.
\end{equation}
We can see that $\mathbf{P}_2$ is always along $y$ direction, i.e. the $\hat{e}_2$ direction. At the same time $\mathbf{P}_1$ is not along the $\hat{e}_1$ direction any more, due to the difference between $\alpha_1$ and $\alpha_3$. This is the very reason why scattering by aligned aspherical grains is different from scattering by spherical grains.

In the third left panel of Figure~\ref{fig:sphere}, we can see that the maximum polarization is still $p=1$, i.e. fully polarized. The location where $p=1$ is achieved is plotted as a solid curve in the figure. The $p=1$ curve for the spherical case is also plotted in the figure as a dashed line. We can see that the $p=1$ locations are slightly different between these two cases. The difference is zero at $\theta=90^\circ$, $\phi=90^\circ$ or $270^\circ$ directions. This is because these two directions correspond to the $\pm \hat{y}$ direction, which is along the dipole $\mathbf{P}_2$. As a result, $\mathbf{P}_2$ does not contribute to the scattered light, and the scattered light is fully polarized. 

\begin{figure}
    \centering
    \includegraphics[width=0.4\textwidth]{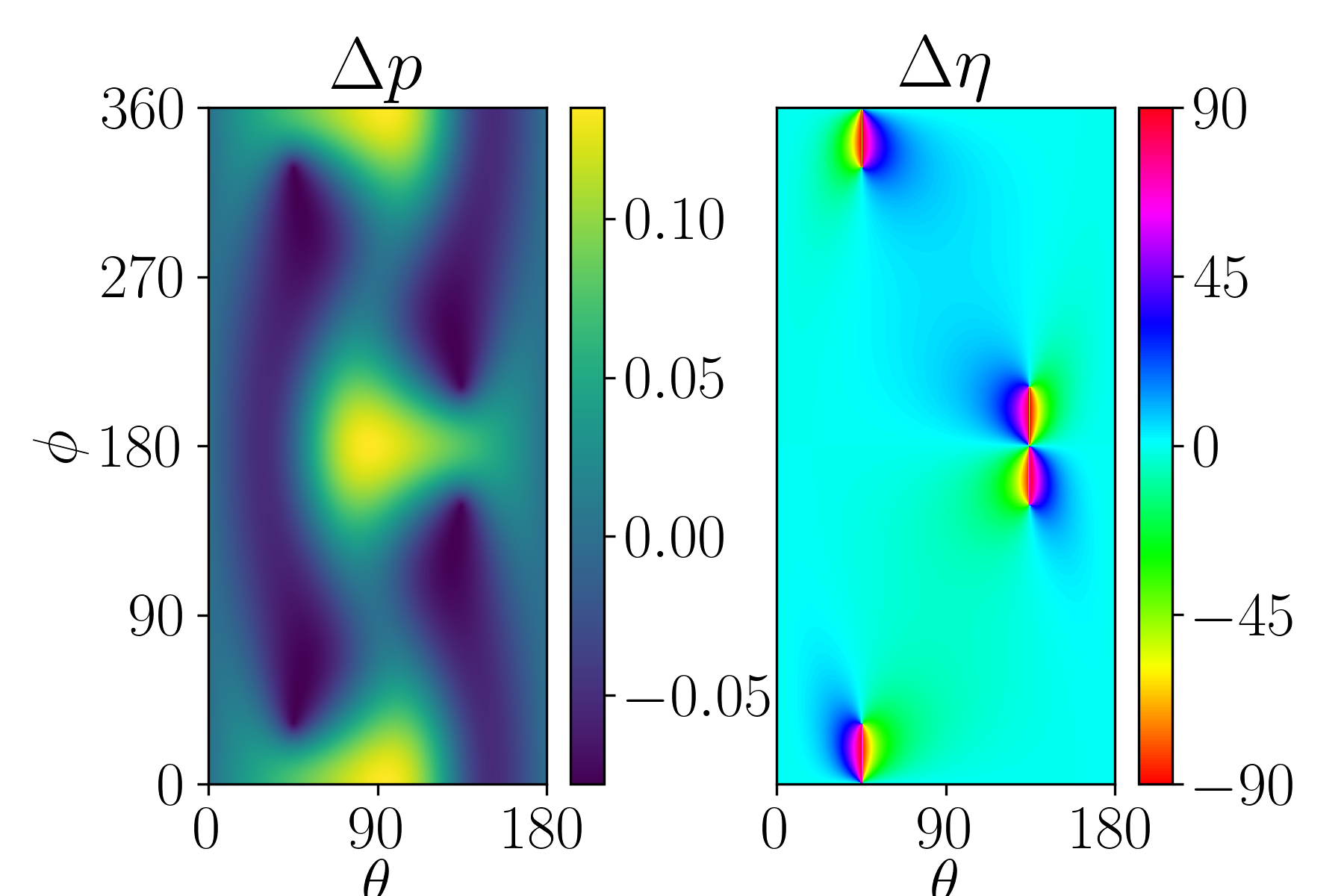}
    \includegraphics[width=0.58\textwidth]{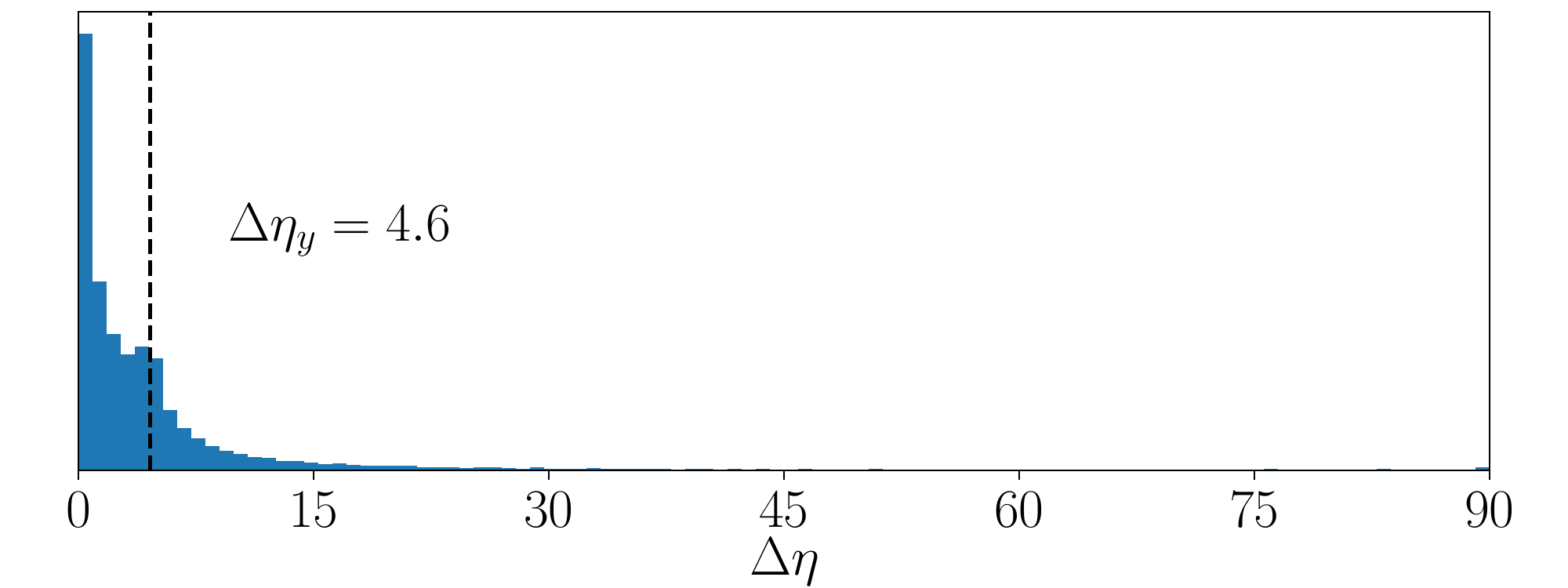}
    \caption{The difference between the spherical and fiducial case. \textit{Left}: The difference in polarization fraction. \textit{Middle}:The difference in polarization orientation. Note that $\Delta\eta=\pm 90^\circ$ denote the same polarization state. \textit{Right}: Histogram of angle difference $\Delta\eta$. The angle difference between the dipole $\mathbf{P}_1$ and $\hat{e}_1$, i.e. $\Delta\eta_y=4.7^\circ$, is also plotted as a vertical dashed line.}
    \label{fig:fid_comp}
\end{figure}

Along the $\phi=0$ line in the right most panel of Figure~\ref{fig:sphere}, the fully polarization ($p=1$) is achieved at $\theta=130.3^\circ$. It differs from $135^\circ$ in the spherical case, by $4.7^\circ$, due to the difference in direction between $\mathbf{P}_1$ and $\hat{e}_1$. This is also the difference in the polarization orientation $\eta$ (Right panel of Figure~\ref{fig:fid_comp}) at $\theta=90^\circ$ and $\phi=90^\circ$ or $270^\circ$, which is for the same reason.

In the right panel of Figure~\ref{fig:sphere}, we first notice that the singular points for the spherical grains ($(\theta=45^\circ, \phi=0^\circ)$ and $(\theta=135^\circ, \phi=180^\circ)$) are no longer singular. There are two reasons for this behavior. Firstly, the forward scattering direction $\hat{n}_1$ is no longer perpendicular to $\mathbf{P}_1$. As a result, the emission is no longer maximized for the dipole radiation from $\mathbf{P}_1$, making it inferior to the radiation from $\mathbf{P}_2$. Secondly and more importantly, $|\mathbf{P}_1|<|\mathbf{P}_2|$ because $\alpha_1>\alpha_3$. These two reasons combine to make the dipole radiation from $\mathbf{P}_2$ dominate over that from $\mathbf{P}_1$ in the forward and backward scattering directions. At these two singular points in the spherical cases, the polarization is thus along $\mathbf{P}_2$ direction, with $\eta=90^\circ$. Because $\eta$ near these points can be $0^\circ$ when varying along constant $\theta$, the difference in polarization orientation between the spherical case and aligned aspherical case is always as large as $90^\circ$ near these points.
There still exist singular points in the diagram for the aspherical case. They are located symmetrically around the previous singular points with the same $\theta$ but different values of $\phi$ between the spherical and fiducial cases.

In the left two panels of Figure~\ref{fig:fid_comp}, we show the difference in polarization degree and the difference in $\eta$ for different scattering directions. We can see that the difference in polarization degree can reach up to $13\%$. The difference in $\eta$ strongly depends on the scattered light direction. Near the forward and backward scattering direction, i.e. the singular points in the spherical case, $\Delta\eta\approx 90^\circ$, which applies up to the new singular points and forms ribbon-like structures in the $\Delta\eta$ plot. 

In the right panel of Figure~\ref{fig:fid_comp}, we show the histogram of the angle difference $\Delta\eta$. We use a vertical dashed line to show $\Delta\eta=4.7^\circ$, which is the angle difference between $\hat{e}_1$ and $\mathrm{P}_1$. We can see that while most scattering directions have $\Delta\eta\lesssim 4.7^\circ$, a fraction of them have substantially larger $\Delta\eta$ of $10^\circ-20^\circ$.

\subsection{Dependence on the dust aspect ratio}
\label{ssec:sGF}

To compare the angle difference with different dust models and/or different inclination angles, we propose the following two metrics. The first one is the $\Delta\eta$ at the scattering angle of $\theta=90^\circ$ and $\phi=90^\circ$, which we call $\Delta\eta_{y}$. This is always the direction along $\mathbf{P}_2$ (and $\hat{y}$) and hence the scattered light comes purely from $\mathbf{P}_1$, so that the $\Delta\eta_{y}$ equals the angle difference between $\mathbf{P}_1$ and $\hat{e}_1$. 

\begin{figure}
    \centering
    \includegraphics[width=0.4\textwidth]{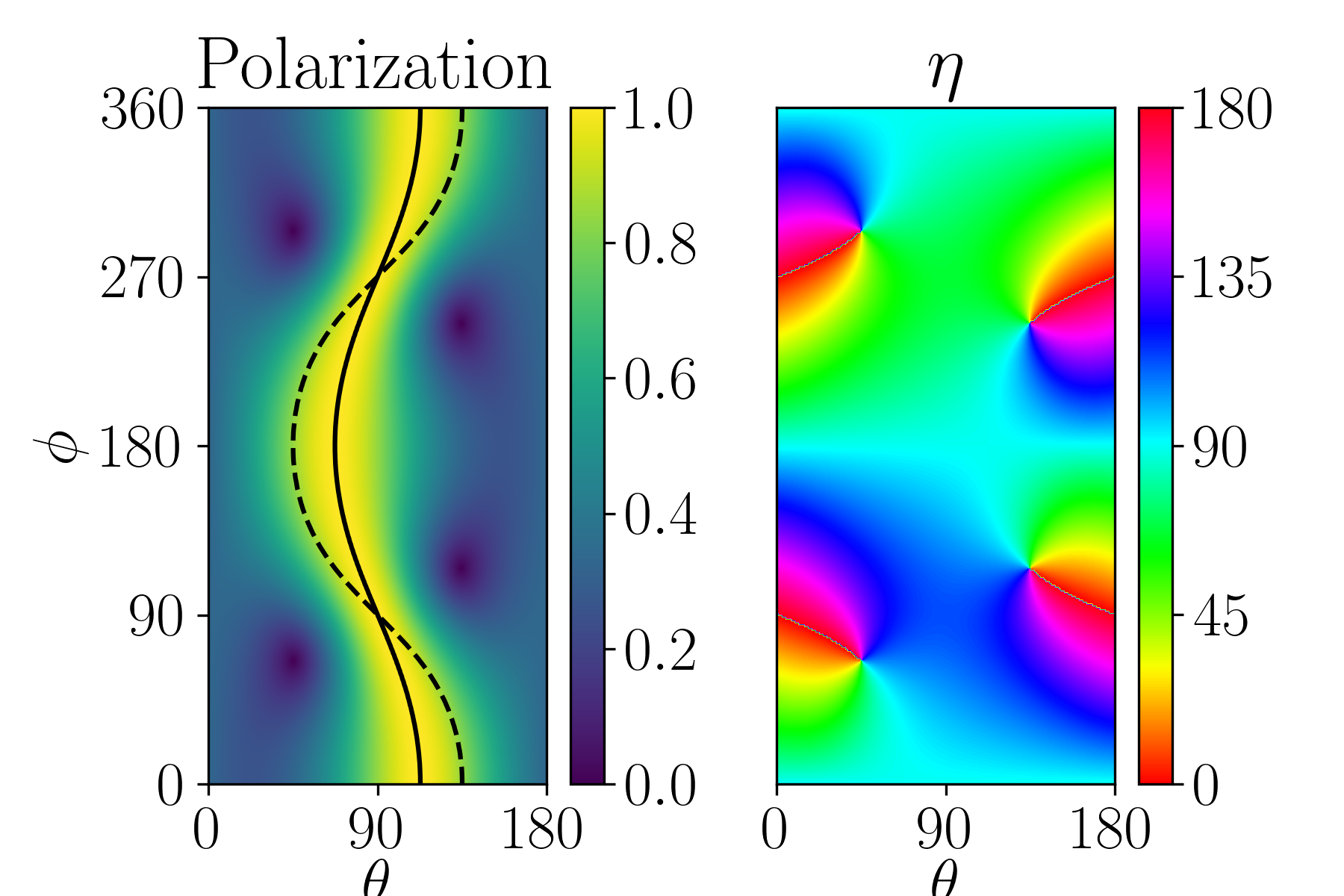}
    \includegraphics[width=0.58\textwidth]{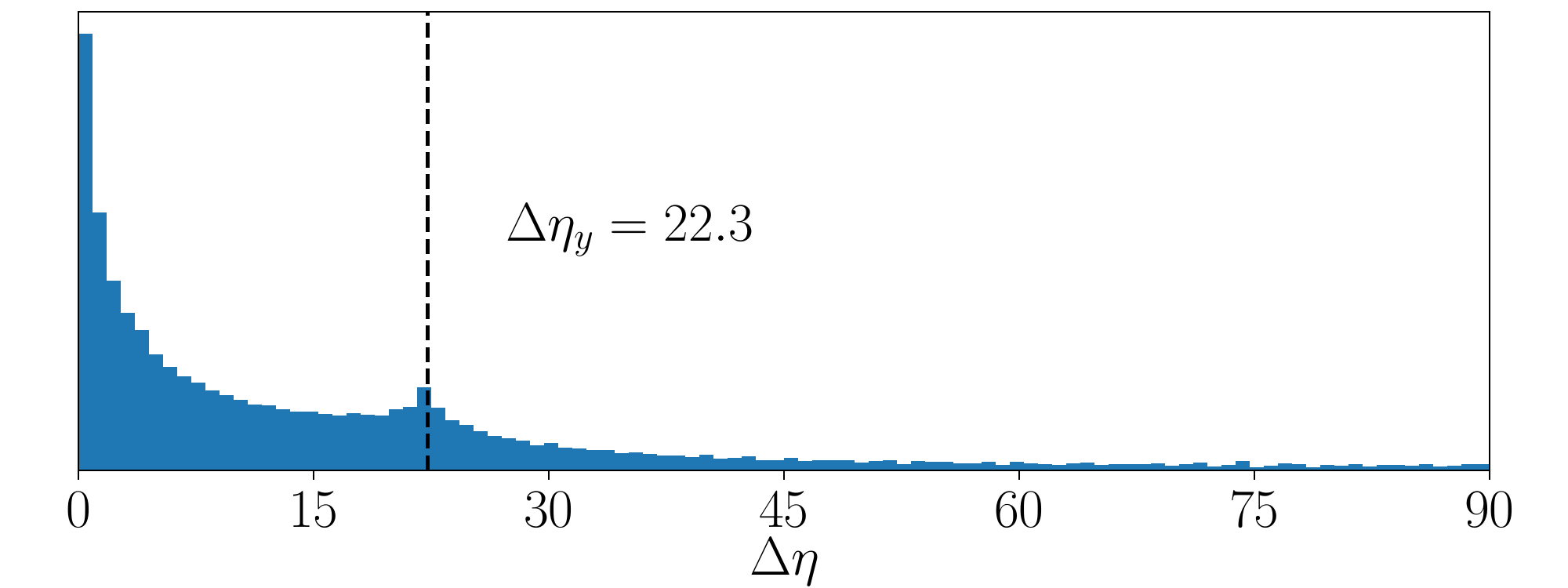}
    \caption{Results for an extremely flattened oblate grain with $s=100$. The left two panels are the same as Figure~\ref{fig:sphere}. The right panel is the same as the right panel of Figure~\ref{fig:fid_comp}.}
    \label{fig:s100}
\end{figure}

In the upper panels of Figure~\ref{fig:s100}, we show the results for an extreme case with $s=100$ and its comparison with the spherical case, again assuming $i=45^\circ$. We can clearly see in the upper left panel that the $p=1$ curve moves closer to $\theta=90^\circ$ line. This is due to the fact that  $|\alpha_1|\gg |\alpha_3|$, so that $\mathbf{P}_1$ is close to $\hat{x}$. In the limit that $\mathbf{P}_1\parallel \hat{x}$, it can be easily verified that only scattered lights in the $xy$ plane, with $\theta=90^\circ$, are fully polarized. 

In the lower panel of Figure~\ref{fig:s100}, we show the histogram for the $s=100$ case. The vertical dashed line represents $\Delta\eta_y=22.3^\circ$. We can see that $\Delta\eta_y$ nicely characterizes the angle difference in this case as well, with most scattered light having an angle difference comparable to or less than $\Delta\eta_y$. 

While $\Delta\eta_y$ sets a scale for the most probable angle difference, it does not provide a good description for the spread in the histogram beyond $\Delta\eta_y$. To characterize the spread, we propose a second metric $\mathrm{max}(p\Delta\eta)$, the maximum value of $p\Delta\eta$ for all scattering directions. It is motivated by the fact that large $\Delta\eta$ directions tend to have low polarization fractions (see Figure~\ref{fig:sphere} and Figure~\ref{fig:s100}). 
In the case of $s=100$, we have $\mathrm{max}(p\Delta\eta)=29.7^\circ$. Note that we always have $\mathrm{max}(p\Delta\eta)\ge \Delta\eta_y$, because when scattered towards $\hat{y}$, we have $p=1$ and $\Delta\eta=\Delta\eta_y$. The difference between $\mathrm{max}(p\Delta\eta)$ and $\Delta\eta_y$ is a measure of the spreading beyond $\Delta\eta_y$.

\begin{figure}
    \centering
    \includegraphics[width=0.49\textwidth]{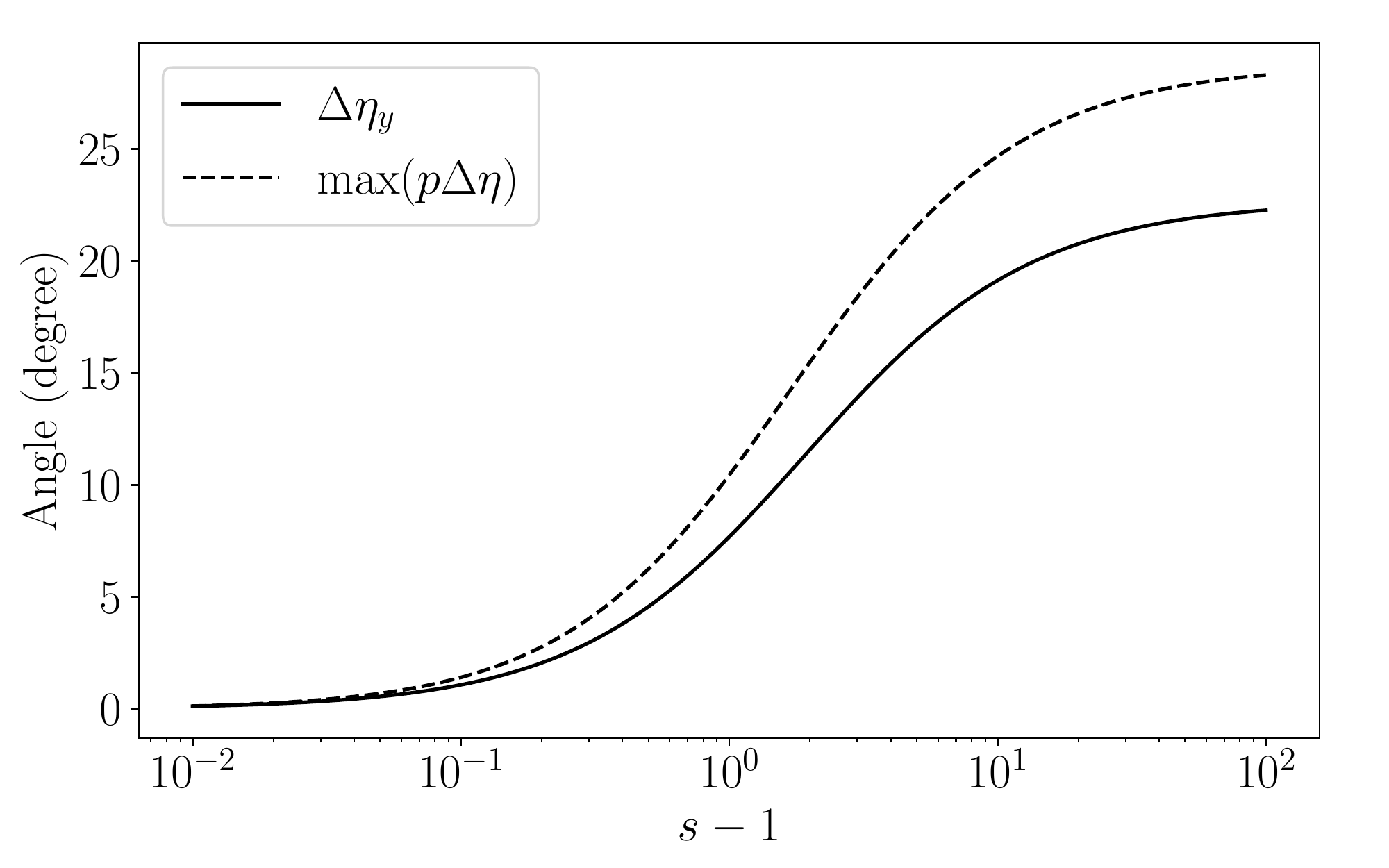}
    \includegraphics[width=0.49\textwidth]{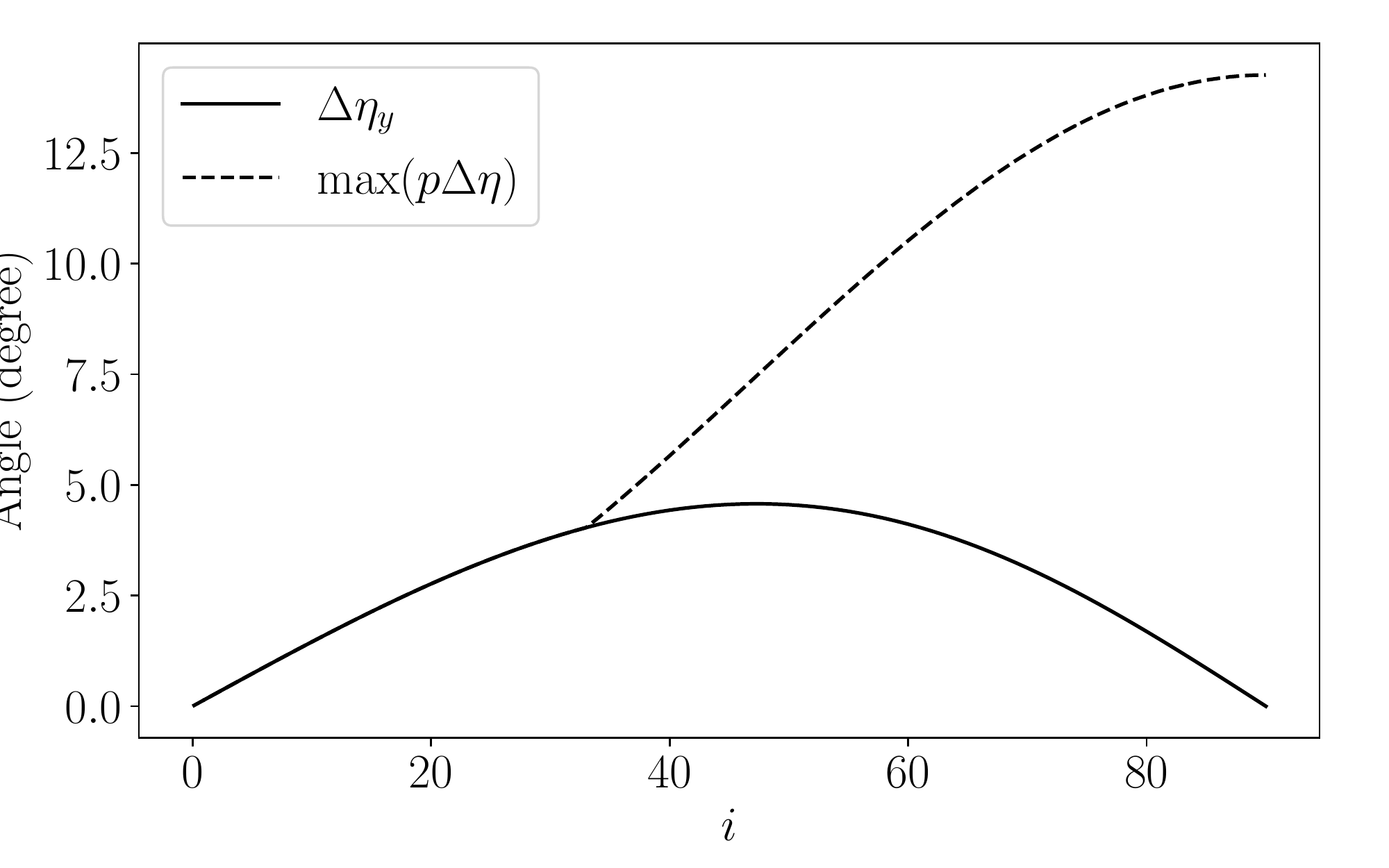}
    \caption{\textit{Left}: The two metrics of the angle difference as a function of the  different dust aspect ratio $s$. 
    \textit{Right}: The two metrics for the angle difference as a function of the incoming light inclination angle $i$ for a fixed grain aspect ratio of $s=1.5$. 
    See text for the definitions of the two metrics.}
    \label{fig:comps}
\end{figure}

In the left panel of Figure~\ref{fig:comps}, we show the two metrics, $\Delta\eta_y$ and $\mathrm{max}(p\Delta\eta)$, as a function of the aspect ratio $s$. We can clearly see that more flattened (oblate) grains have larger deviations in polarization orientation due to scattering compared to spherical grains. For $s=2$, the angle difference is typically on order of $10^\circ$, while for extremely elongated grains, the difference can be as large as $30^\circ$ for a large fraction of scattering angles. 

\subsection{Dependence on the inclination angle}
\label{ssec:incGF}

In the right panel of Figure~\ref{fig:comps}, we show the two metrics, $\Delta\eta_y$ and $\mathrm{max}(p\Delta\eta)$, as a function of the inclination angle $i$. We can see that $\Delta\eta_y$ increases as we increase the inclination angle $i$ initially, then falls back to $0$ as we go towards $i=90^\circ$. In the two limiting cases, $i=0$ and $i=90^\circ$, we have $\Delta\eta_y=0$. This is because $\hat{e}_1$ becomes aligned with one of the principle axes ($\hat{z}$ for $i=0$ and $\hat{x}$ for $i=90^\circ$), hence $\mathbf{P}_1\parallel\hat{e}_1$. However, this doesn't imply the angle difference also goes to $0$ as the inclination angle approaches $90^\circ$. In fact, for larger inclination angles, the difference between $|\mathbf{P}_1|$ and $|\mathbf{P}_2|$ increases with $i$, and the polarization orientation deviates from the spherical case (with $|\mathbf{P}_1|=|\mathbf{P}_2|$) progressively.
These differences are not captured by the first metric $\Delta\eta_y$.

\begin{figure*}
    \includegraphics[width=0.4\textwidth]{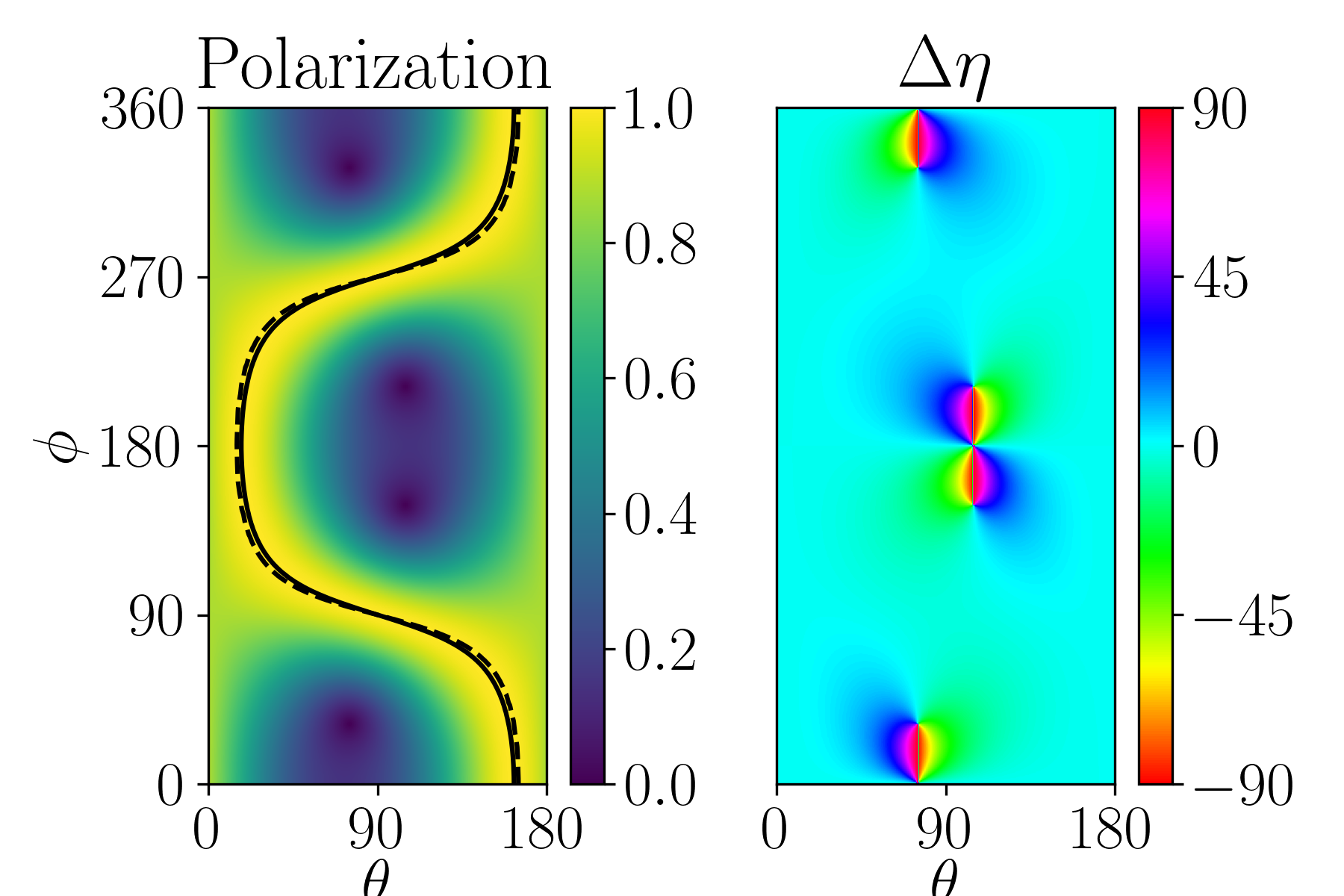} 
    \includegraphics[width=0.58\textwidth]{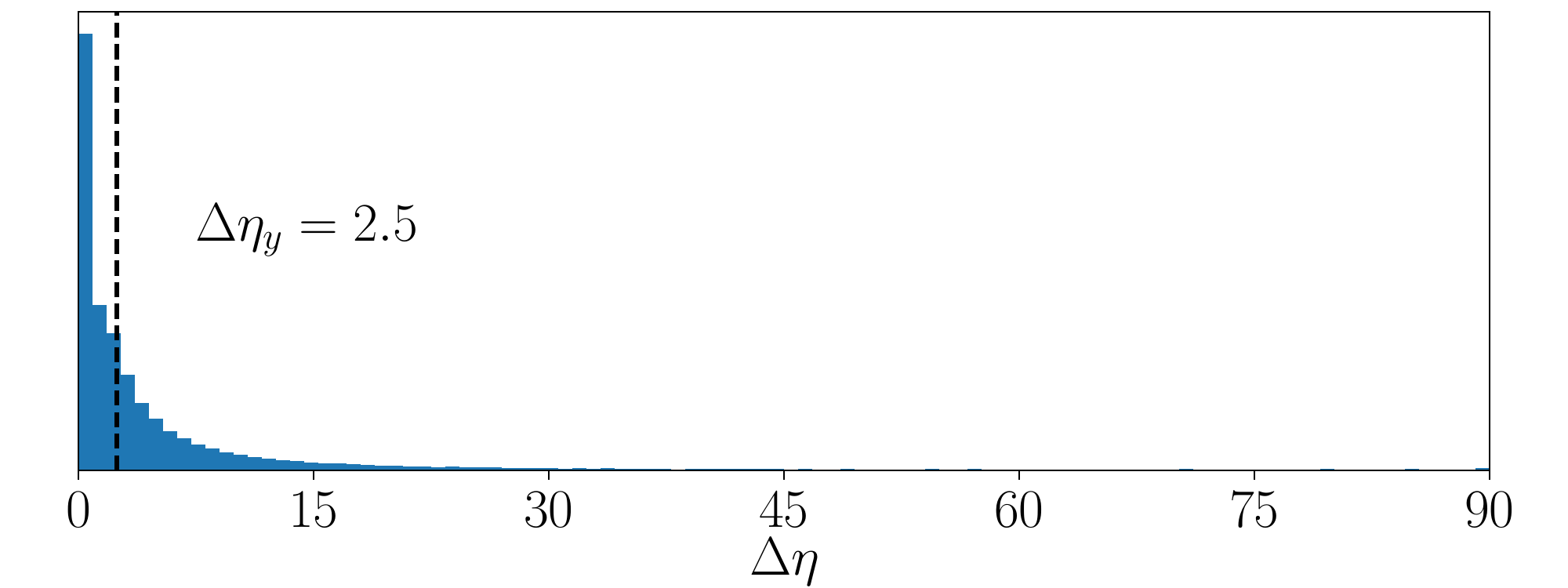} 
    \includegraphics[width=0.4\textwidth]{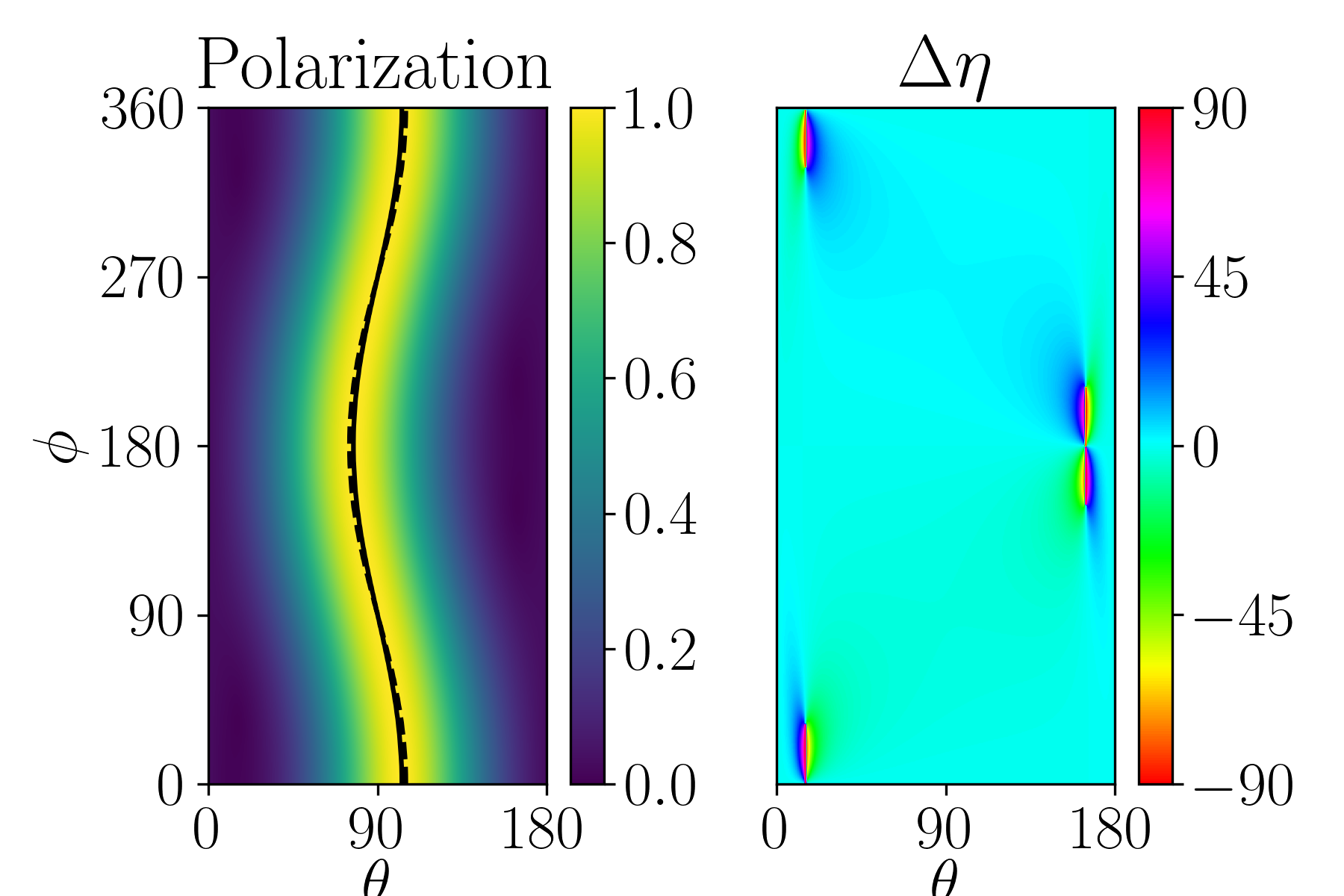} 
    \includegraphics[width=0.58\textwidth]{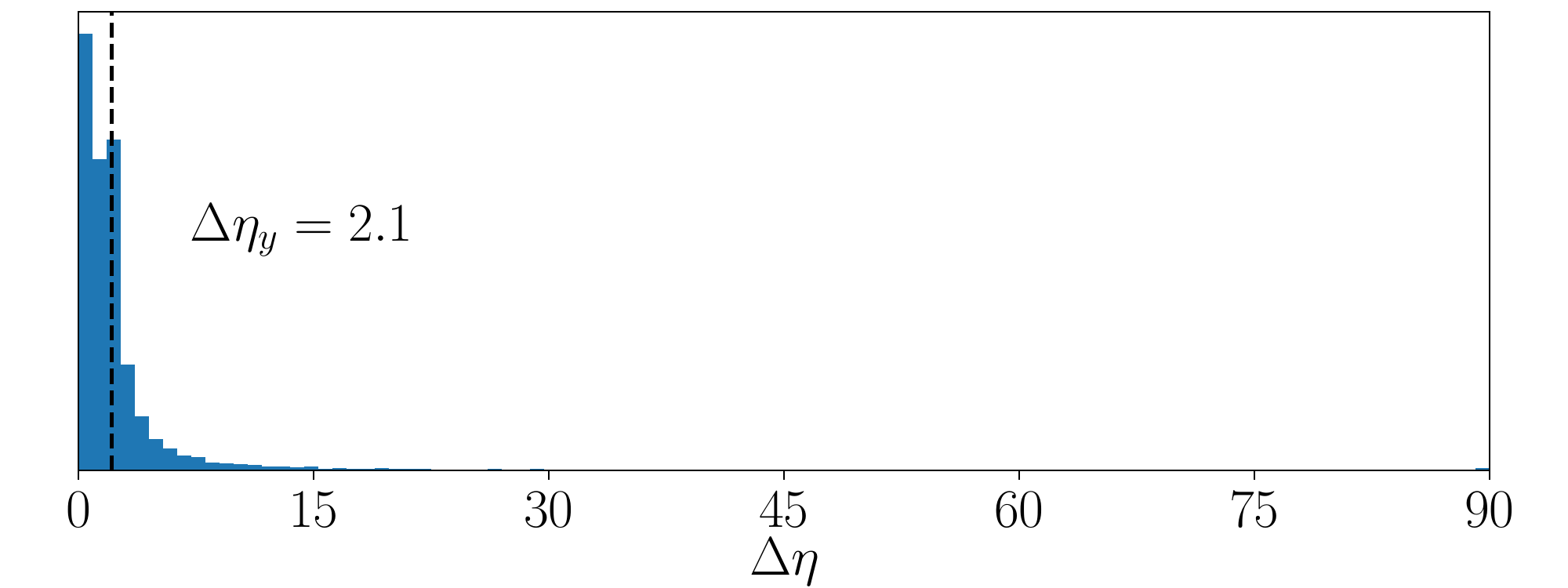}
\caption{Results for $i=75^\circ$ and $i=15^\circ$. \textit{Top panels:} results for $i=75^\circ$. From left to right shows the polarization fraction, the angle difference with spherical case, and the histogram of angle differences. \textit{Bottom panels:} results for $i=15^\circ$ in the same order.}
\label{fig:i75v15}
\end{figure*}

At a larger inclination angle, there is a stronger need for the second metric. To see what exactly happens at a larger inclination angle, we show the polarization fraction and angle difference for $i=75^\circ$ in upper panels of Figure~\ref{fig:i75v15}. To compare, we show the same for $i=15^\circ$ in lower panels of Figure~\ref{fig:i75v15}. These two models have similar $\Delta\eta_y$ ($2.5^\circ$ and $2.1^\circ$ for $i=75^\circ$ and $i=15^\circ$, respectively). We can see that even though the deviations of the $p=1$ locations from corresponding spherical models are similar for these two inclination angles, the ribbon-like structures are much larger in $i=75^\circ$ than in $i=15^\circ$. This results in a larger spread in the histogram of angle differences in $i=75^\circ$ than $i=15^\circ$, shown in the right panels of Figure~\ref{fig:i75v15}.

This spread is nicely captured by our second metric, $\mathrm{max}(p\Delta\eta)$, which are $13.2^\circ$ and $2.1^\circ$ at $i=75^\circ$ and $i=15^\circ$, respectively. According to $\mathrm{max}(p\Delta\eta)$, there is a substantial fraction of scattering directions with angle differences as large as $\sim 13^\circ$ when $i=75^\circ$, even though the $\Delta\eta_y$ is only $2.5^\circ$. 

With a better understanding of the two metrics, $\Delta\eta_y$ and $\mathrm{max}(p\Delta\eta)$, we now come back to the right panel of Figure~\ref{fig:comps}. We conclude that the angle difference with the spherical model increases monotonically as the inclination angle $i$ increases. There is a large fraction of scattering directions with $\Delta\eta$ of $\sim 15^\circ$ or larger when $i$ is close to $90^\circ$. 

Interestingly, $\mathrm{max}(p\Delta\eta)=\Delta\eta_y$ for small inclination angles. In these cases, the ribbon-like structures are very thin and the angle difference is dominated by the difference between $\mathrm{P}_1$ and $\hat{e}_1$ (see upper lower panels of Figure~\ref{fig:i75v15}).

\subsection{Dependence on dust composition}
\label{ssec:composition}

The composition of dust has a strong impact on the optical properties of dust grains. To show the dependence of angle difference on dust composition, we calculated $\Delta\eta_y$ and $\mathrm{max}(p\Delta\eta)$ assuming $i=45^\circ$ and $s=1.5$ for five illustrative dust compositions. The results are tabulated in Table~\ref{tab:composition}. We also listed the real ($n$) and the imaginary part ($k$) of the refractive index for each composition at the wavelength $\lambda=1.5\micron$.

\begin{table}
    \begin{center}
    \begin{tabular}{|c|cc|cc|}
        \hline                 
        Composition &  $n$   &  $k$   & $\Delta\eta_y$ & $\mathrm{max}(p\Delta\eta)$ \\
        \hline                 
        DSHARP$^a$      & $1.56$ & $2.0\times 10^{-2}$ & $4.6^\circ$  & $6.2^\circ$ \\
        Silicate$^b$    & $1.69$ & $3.2\times 10^{-2}$ & $5.3^\circ$  & $7.3^\circ$ \\
        Troilite$^c$    & $6.57$ & $2.59$              & $12.6^\circ$ & $16.8^\circ$ \\
        Water ice$^d$   & $1.29$ & $4.7\times 10^{-4}$ & $2.6^\circ$  & $3.5^\circ$ \\
        Organics$^c$    & $1.62$ & $2.0\times 10^{-2}$ & $4.9^\circ$  & $6.8^\circ$ \\
        \hline                 
    \end{tabular}
    \caption{Angle difference for different composition assuming $i=45^\circ$ and $s=1.5$ at an observing wavelength of $1.5\rm\, \mu m$. }
    \end{center}
    
    $^a$: \cite{Birnstiel2018}; 
    
    $^b$: \cite{Draine2003}; 
    
    $^c$: \cite{Henning1996}; 
    
    $^d$: \cite{Warren2008}.
    \label{tab:composition}
\end{table}

Among the materials considered, troilite produces the largest angle difference. This may be related to its absorptive nature, characterized by its large imaginary part of the refractive index $k$.

\subsection{Results for moderately large dust grains}
\label{ssec:large}

\begin{figure}
    \centering
    \includegraphics[width=0.4\textwidth]{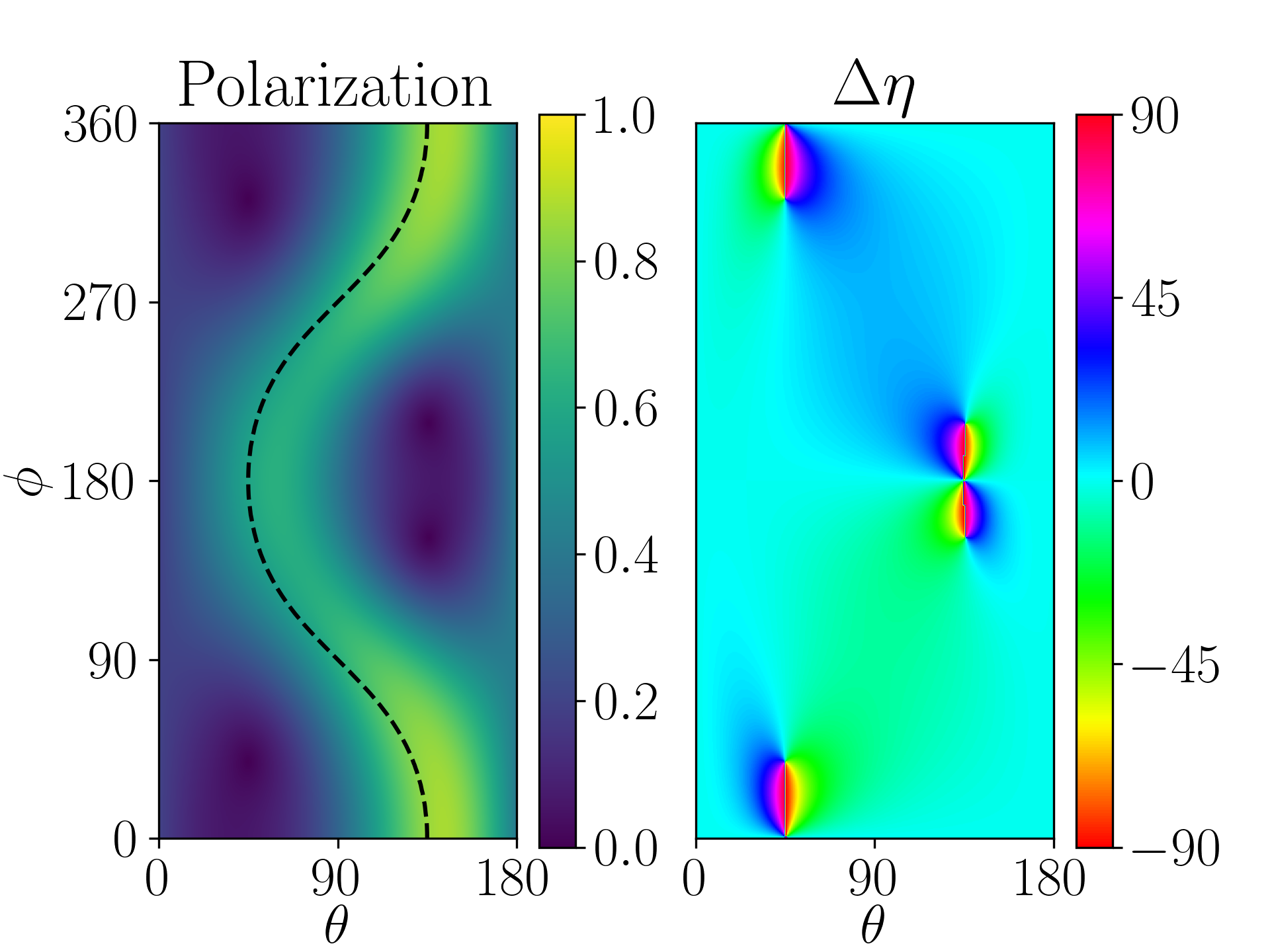}
    \includegraphics[width=0.58\textwidth]{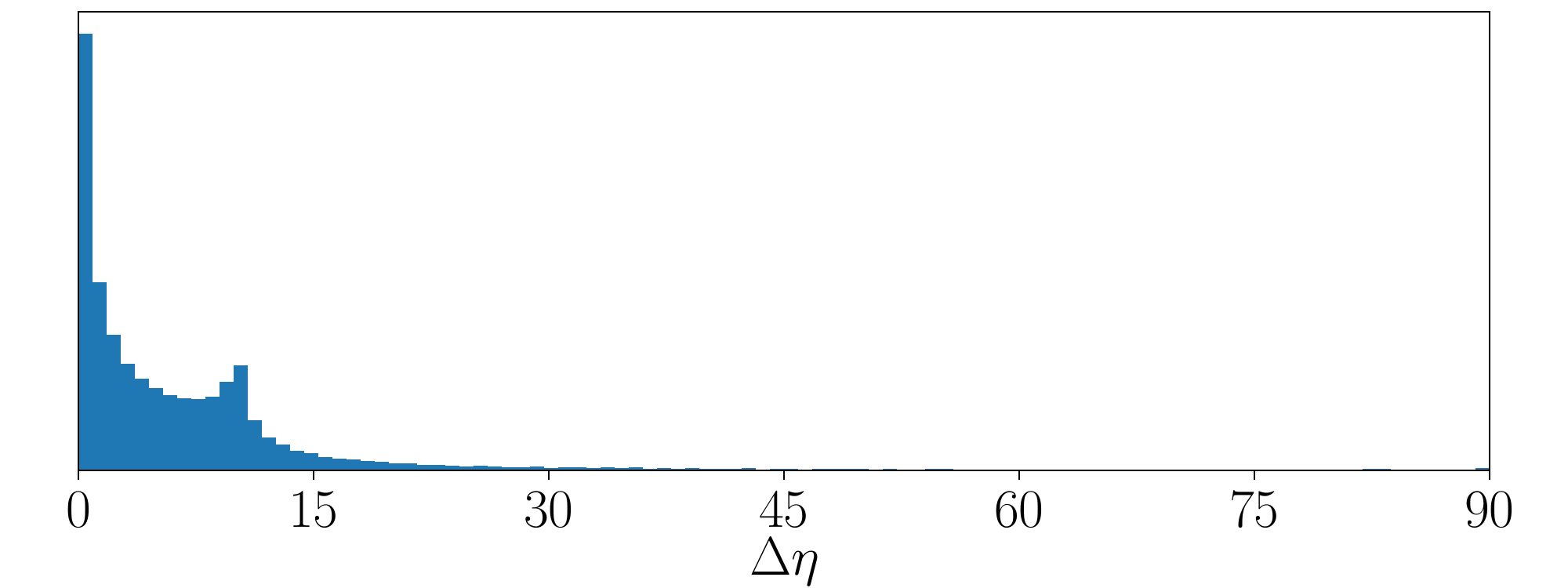}
    \caption{The same as Figure~\ref{fig:i75v15} but for an MRN-distributed dust grains with $a_\mathrm{max}=1\micron$.}
    \label{fig:mrn}
\end{figure}

The size of grains has a strong impact on dust scattering. To relax the previous small grain size assumption, we use the PyTMatrix\footnote{Available at \url{https://github.com/jleinonen/pytmatrix/}} module \citep{Leinonen2014}, which is a wrapper for the TMatrix code \citep{Mishchenko1994}. 
The results for MRN-distributed dust grains with $s=1.5$ and $a_\mathrm{max}=1\micron$, corresponding to a size parameter of $x_\mathrm{max}=4.2$, are shown in Figure~\ref{fig:mrn}. 
We can see that the polarization fraction decreases significantly from $100\%$ at the peak curve, but the distribution of $\eta$ is still similar to the one in dipole regime in right panels of Figure~\ref{fig:sphere}. The histogram of the angle difference  is also larger for the $1\micron$ grains. The dipole approximation is reasonable for $1\micron$ grains or smaller, if we focus on the polarization orientation $\eta$. 

We note that even larger grains ($a_\mathrm{max}\geq 2\micron$) can produce more complicated polarization patterns and distributions of $\eta$ (results not presented in this paper), which are harder to use for interpreting observational results. Larger grains may account for the strong forward scattering observed in some systems (e.g. \citealt{Avenhaus2018}). We will postpone a full exploration of larger dust grains to a future investigation and focus on the dipole approximation in this work.

\section{Dipole approximation}
\label{apsec:dipole}

Since the dipole approximation is an important part of our methodology, we will describe it briefly together with several key equations to help understand the results in this paper, especially the importance of the quantity $\Delta \eta_y$ defined in Section~\ref{sec:GF}. In particular,  Equation~\ref{eq:Smatrix} is the key to calculate the amplitude scattering matrix but is not in the literature as far as we know. For a more detailed derivation and description, we refer interested readers to \cite{BH83}, especially their Section~5. 

When the dust grains are small compared to the observing wavelength, the response of the particle to an external electromagnetic wave can be well represented by a dipole oscillating with the same frequency and phase as the incoming radiation. Let $\mathbf{E}_i$ be the E vector of the incoming light. The excited dipole is linear with respect to $\mathbf{E}_i$: $\mathbf{P}=\alpha\mathbf{E}_i$, where $\alpha$ is the $3\times 3$ polarizability matrix of the dust grain. 

The oscillating dipole $\mathbf{P}$ will then radiate a secondary electromagnetic wave with an electric field (in far field with $kr\gg1$):
\begin{equation}
\mathbf{E}_s = \frac{k^2}{r}[(\hat{r}\times \mathbf{P})\times \hat{r}]e^{i(kr-\omega t)},
\label{eq:Es}
\end{equation}
where $\hat{r}\equiv \mathbf{r}/r$. Decompose the incoming radiation as $\mathbf{E}_i=E_{i1}\hat{e}_{i1}+E_{i2}\hat{e}_{i2}$, the scattered radiation as $\mathbf{E}_s=E_{s1}\hat{e}_{s1}+E_{s2}\hat{e}_{s2}$, we can define the following amplitude scattering matrix:
\begin{equation}
\left(\begin{array}{c}E_{s1}\\ E_{s2}\end{array}\right) = 
\frac{e^{ik(r-z)}}{-ikr}
\left(\begin{array}{cc}S_{11}& S_{12}\\ S_{21} & S_{22}\end{array}\right)
\left(\begin{array}{c}E_{i1}\\ E_{i2}\end{array}\right),
\end{equation}
thanks to the far-field dependence of scattered light in Equation~\eqref{eq:Es}.
Calculating the dot product of $\hat{e}_{s1}\cdot \mathbf{E}_s$ from Equation~\eqref{eq:Es}, we can easily derive the $S_{11}$ and $S_{12}$. Similarly, we can derive the $S_{21}$ and $S_{22}$ through $\hat{e}_{s2}\cdot \mathbf{E}_s$. They can be nicely summarized as follows:
\begin{equation}
S_{mn} = (-ik^3)(\hat{e}_{sm}\cdot\alpha\cdot\hat{e}_{in}),
\label{eq:Smatrix}
\end{equation}
with $m,n=1,2$. 
Equation~\eqref{eq:Smatrix} is rotation invariant, and can be evaluated in any frame. The dipole approximation is thus free of rotation of Stokes parameters: one can rotate the polarizability matrix into the Disk Frame once and for all instead of rotating Stokes parameters into the Grain's Frame for every scattering event.

With the amplitude scattering matrix calculated in Equation~\eqref{eq:Smatrix}, the Mueller Matrix relating the incoming and scattering Stokes parameters can be calculated using Equation (3.16) of \cite{BH83}.

In this work, we use oblate spheroidal particles to represent aligned elongated dust grains. In this case, let $a_1=a_2>a_3$ be the three principle semi-major axes and $a^3=a_1a_2a_3$ be the effective radius of the dust grain. We have a diagonal matrix for the polarizability matrix: $\alpha = \mathrm{diag}\left\{\alpha_1,\alpha_1,\alpha_3\right\}$, with $|\alpha_1|>|\alpha_3|$. The polarizability is:
\begin{equation}
\alpha_l = a^3 \frac{\epsilon-1}{3+2L_l(\epsilon-1)},
\label{eq:polmatrix}
\end{equation}
where $\epsilon$ is the complex dielectric function, $L_l (l=1,2,3$) are geometric factors with $L_1+L_2+L_3=1$. For a oblate spheroid, we have 
\begin{equation}
L_1 = \frac{g(e)}{2e^2}\left[\frac{\pi}{2}-\mathrm{tan}^{-1}g(e)\right]-\frac{g^2(e)}{2},
\end{equation}
where $e^2=1-a_3^2/a_1^2$ is the eccentricity, not to be confused with natural base $e$ in other context. Function $g(e)\equiv \sqrt{(1-e^2)/e^2}$, and $L_3=1-2L_1$.

\end{CJK*}
\end{document}